%% file: RT-0323.tex
\documentclass[a4paper,english]{article}
\usepackage{RR}
\RRNo{0323}
\usepackage{hyperref}
\usepackage{epsfig}
\usepackage{graphics}
\usepackage[english]{babel}
\usepackage{moreverb}

\RRdate{June 2006}
\RRauthor{
Pierre Parrend\and
Stéphane Frénot
  \thanks[sfn]{This Work is partialy founded by Muse IST Project n°026442.}
\thanksref{sfn}
}
\authorhead{Parrend \& Frénot}
\RRtitle{Déploiement sécurisé de composants pour la plate-forme OSGi$^{tm}$ Release 4}
\RRetitle{Secure Component Deployment in the OSGi$^{tm}$ Release 4 Platform}

\titlehead{Secure Component Deployment}
\RRresume{L'utilisation de plates-formes à composants a connu ces dernières années une forte croissance, en particulier dans le domaine des Systèmes Embarqués. La plate-forme OSGi$^{tm}$ est une de ces plates-formes dédiées aux environnements légers. Cependant, la mise en oeuvre de nouvelles plates-formes implique la présence de nouveaux risques de sécurité, ainsi qu'un manque à la fois de connaissance et d'outils pour palier à ces nouveaux risques.

Ce rapport technique a pour objectif d'améliorer la compréhension des mécanismes de sécurité dans le cadre du déploiement de composants. Il se concentre sur la problématique de la sécurisation du déploiement. Il présente les mécanismes de cryptographie nécessaires à la signature des composants OSGi$^{tm}$ (appelés bundles), et le processus détaillé de signature et de validation de ces bundles.

Nous présentons également la plate-forme SFelix, qui est une implémentation sûre du framework OSGi$^{tm}$ basée sur le projet Apache Felix. SFelix comprend notre implémentation du processus de validation de bundles, conforme à la spécification OSGi$^{tm}$ release 4. De plus, un outil de signature et de publication de bundle SFelix JarSigner, a été développé de manière à intégrer la sécurisation des bundles dans le processus de déploiement.}

\RRabstract{Last years have seen a dramatic increase in the use of component platforms, not only in classical application servers, but also more and more in the domain of Embedded Systems. The OSGi$^{tm}$ platform is one of these platforms dedicated to lightweight execution environments, and one of the most prominent. However, new platforms also imply new security flaws, and a lack of both knowledge and tools for protecting the exposed systems.

This technical report aims at fostering the understanding of security mechanisms in component deployment. It focuses on securing the deployment of components. It presents the cryptographic mechanisms necessary for signing  OSGi$^{tm}$ bundles, as well as the detailed process of bundle signature and validation. 

We also present the SFelix platform, which is a secure extension to Felix OSGi$^{tm}$ framework implementation. It includes our implementation of the bundle signature process, as specified by OSGi$^{tm}$ Release 4 Security Layer. Moreover, a tool for signing and publishing bundles, SFelix JarSigner, has been developed to conveniently integrate bundle signature in the bundle deployment process.}
\RRmotcle{OSGi$^{tm}$, Sécurité, Intégrité, Authentification, Signature de Jar, Signature Numérique, Felix.}
\RRkeyword{OSGi$^{tm}$, Security, Integrity, Authentication, Jar Signature, Digital Signature, Felix.}

\RRprojet{ARES}
\RRtheme{\THCom} % cas d'un seul theme

\URRhoneAlpes % pour ceux qui sont dans les montagnes

\begin{document}

\makeRT % cas d'un rapport technique.

\tableofcontents{}

\newpage
\listoffigures

\newpage
\listoftables

\newpage
\input{introduction}

\newpage
\input{cryptography}

\newpage
\input{secureDeployment}

\newpage
\input{implementation}

\newpage
\input{conclusions}

\newpage
\input{annexes}

\newpage
\bibliographystyle{alpha}
\bibliography{rt6-signing}

\end{document}

%% file: introduction.tex
\section{Introduction}
\label{introduction}

\subsection{Context of this Report}

The OSGi$^{tm}$ platform is an execution layer over the Java Virtual Machine that supports life cycle management of components (introduction, update, removal) during runtime. These components provide Java packages or locally published services (as Java interfaces) to other components.

This technical report aims at fostering the understanding of security mechanisms in the OSGi$^{tm}$ platform. It focuses on securing the deployment of components. It presents the cryptographic mechanisms necessary for signing OSGi$^{tm}$ components (also named bundles), as well as the detailed process of bundle signature and validation. 

We present the SFelix platform\footnote{http://sfelix.gforge.inria.fr}, which is a secure extension to Apache Felix\footnote{http://incubator.apache.org/felix/} OSGi$^{tm}$ framework implementation. It includes our implementation of the bundle validation process in OSGi$^{tm}$ Release 4 Security Layer. Moreover, a tool for signing and publishing bundles, SFelix JarSigner, has been developed to conveniently integrate bundle signature in the bundle deployment process.

\subsection{Component Deployment}

\begin{figure}[htb]
\centering
\epsfig{file=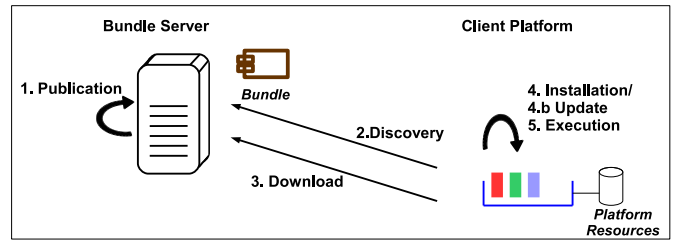, width=250pt}
\caption{The Component Deployment Process}
\label{fig:deployment}
\end{figure}

The deployment of bundles is not defined by the OSGi$^{tm}$ specifications. In the Felix implementation, it is realized by the publication of the bundles on a server on the Internet, and the installation of these bundles from the server by the client platforms. The steps of the deployment process are the following: publication (1), bundle discovery (2), download (3), installation (4) and update (4.b), execution (5). The figure \ref{fig:deployment} shows this process of bundle deployment.

\subsection{Threats during Deployment}

Major security threats during deployment are of three types. The first type is the presence of malicious bundle publication servers.
The second type of deployment threats is man-in-the-middle attack. Such an attacker can modify the bundle, or fully substitute the loaded bundle by another one. In both cases the client platform installs then executes some code without being able to do any assumption about the code quality or reliability. The third type of threats is the possibility that exists for an attacker to access and modify (=to tamper) the stored data used by the component platform. Actually, all configuration data and installed components are available on the filesystem of the host, and access to this filesystem is sufficient to fully control the behavior of the platform and of the components.

So as to protect the component platform from the first two threats, it is necessary to control that the bundle publishers are trustworthy, and that loaded bundles have not been tampered with during the transfer over an untrusted network, such as the Internet. Jar specifications (bundles are specific Jar archives) propose to sign archive so as to guarantee such properties. OSGi$^{tm}$ specifications propose additional restrictions to signing, notably by forbidding uncomplete archive signing, which allows a third party to add resources to an archive without invalidating the signature.

The protection of the platform resources requires to integrate a secure filesystem with the component platform. Bundle substitution or configuration tampering between download and starting is then prevented. As far as it does not deal directly with the problem of securing the deployment process, this extension of the platform will not be considered further in this report.

\begin{figure}[htb]
\centering
\epsfig{file=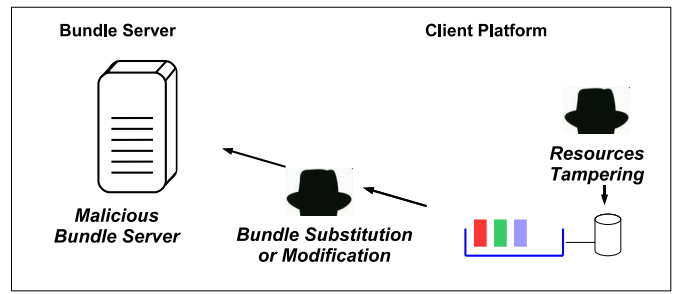, width=250pt}
\caption{The Security Threats during the Component Deployment Process}
\label{fig:deployment-pbs}
\end{figure}

Figure \ref{fig:deployment-pbs} shows the security threats that exist during the deployment process of a component. 

\subsection{Malicious Bundles}

The malicious bundles can be categorized according to the target of their attacks, and according to the attack paradigm they use, that is to say the type of event that triggers the attack.

\paragraph{Attack Targets}
Malicious bundles can be classified in four main types according to the target of the attack they perform:
\begin{enumerate}
 \item \emph{Threat to the System}, for instance, a bundle can contain JNI calls which makes it possible to access the underlying Operating System; or it can have extremely resource intensive services which consume most of the available CPU or memory of the system,
 \item \emph{Threat to the Platform}, a bundle can try and access the Java platform (through the System or Runtime classes), or the OSGi$^{tm}$ platform,
 \item \emph{Threat to the Bundles}, a bundle can misuse available services (depending of the service API), or provide false services, that provides a given service with an improper implementation,
 \item \emph{Undue Monitoring}, a bundle can gather data related to a platform without making immediate use of them, and send them to a remote attacker for latter intrusion.
\end{enumerate}

\paragraph{Attack paradigms} Three main attack paradigms can be used by malicious bundles. They are characterized by the event that triggers the attack:
\begin{enumerate}
 \item \emph{Back doors}. Those bundles make it possible for a remote attacker to gain access to the execution platform.
 \item \emph{Malicious services}. those bundles provide fake services or classes that can be used instead of valid ones.
 \item \emph{Autonomous bundles}. those bundles perform malicious actions autonomously, without remote control and without requiring service calls from other services. Their behavior is close to the one of viruses.
\end{enumerate}

It is of course possible that a malicious bundle uses several of these attack paradigms simultaneously.

This brief presentation of the attacks that may occur through malicious OSGi$^{tm}$ bundles highlights the need for securing the life cycle of the bundles, and particularly for protecting the deployment phase from malicious Bundle Repositories, from bundle substitution during transfer or from local tampering during the installation phase.

\bigskip

This technical report presents the mechanisms that are necessary to implement the bundle signature and validation process. First, underlying cryptographic concepts are presented. Then, the algorithms for signing and validating an OSGi$^{tm}$ bundle are detailed. And, lastly, our implementation of bundle signature process is presented. It is made of the SFelix platform - an extension of Apache Felix OSGi$^{tm}$ implementation- on the first hand, and of SFelix JarSigner tool - that supports signing and publication of bundles-  on the other hand.

%% file: cryptography.tex
\section{Cryptographic Concepts and Standards}
\label{cryptography}

Security of systems is the ability of these systems to withstand the behavior of malicious users. It can be defined as the conjunction of integrity, availability for authorized users only, and confidentiality \cite{avizienis00dependability}. Identification of authorized users is named authentication.

Secure bundle deployment means that these requirements are guaranteed during the whole deployment process, that is to say from the publication of a bundle until the time when this bundle is started. It is based on asymmetric, or public key, cryptography, which publicly binds a given key pair with a unique user. This pair is made of a secret private key owned by the user and of a public key that is widely available. The private key is used to encrypt data. Every third party user can then assert that data that are decryptable with the public key have been encrypted with the private one.

For each of the security requirements that have been defined, a definition will be given, the cryptographic mechanisms necessary for their enforcement will be presented, and supporting standards will be introduced.

 \subsection{Integrity}
\label{integrity}
%  \subsubsection{Definition}

\emph{Integrity} is defined by \cite{avizienis00dependability} as the absence of improper system state alterations. In the deployment process, this means that loaded bundles must not be modified between the publication step and the start step.

  \subsubsection{Digital Signature}

A `Digital Signature' is an electronic signature that can be used to ensure that the original content of the message or document that has been sent is unchanged, and to authenticate the identity of the sender of a message or the signer of a document. That is to say a digital signature guarantees both the integrity of the document and the authentication of its emitter. We will concentrate in these section on integrity. Authentication will be presented in details in section \ref{authentication}.

\begin{figure}[htb]
\centering
\epsfig{file=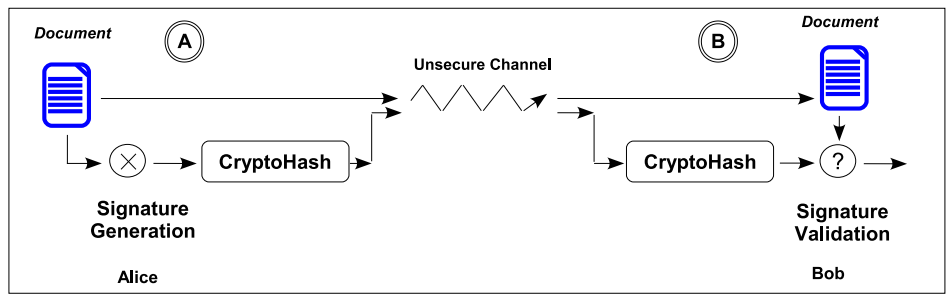, width=360pt}
\caption{Digital Signature of a Document using asymmetric Cryptography}
\label{fig:digital-signature}
\end{figure}

The overview of the process of digital signature is shown in figure \ref{fig:digital-signature}.

 This process can be shared in two separate steps: the signature generation (A) by the emitter of the signed document, and the signature validation (B), by the receiver. Signature generation consists in applying a signature algorithm on the signed document. This algorithm results in a data file called Cryptohash, or more frequently Digital Signature. This signature is passed over from the emitter to the receiver along with the signed document. The receiver can then check whether the digital signature matches the document.

\begin{figure}[h]
\hfill
\begin{minipage}[t]{.45\textwidth}
    \begin{center} 
\epsfig{file=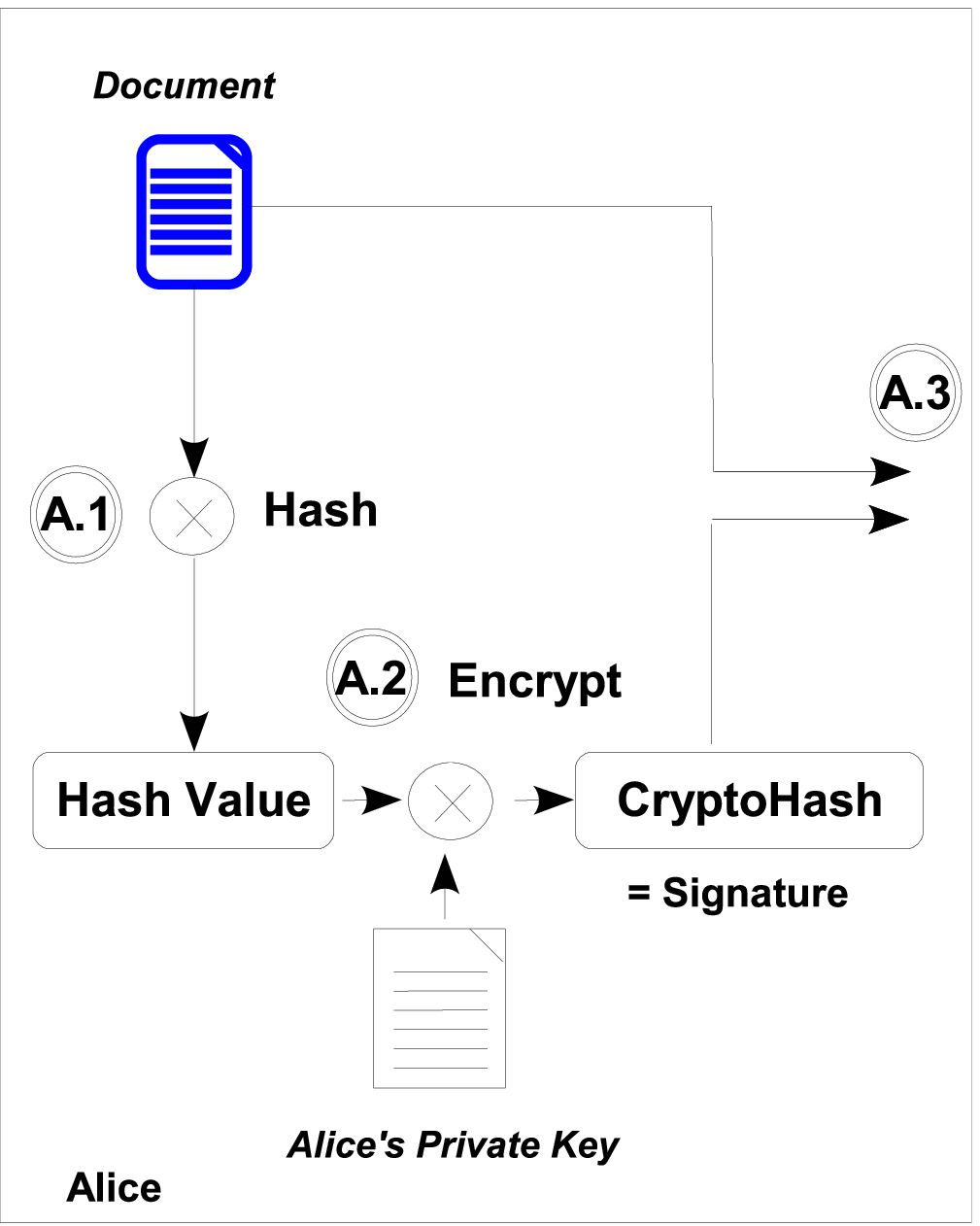, width=130pt}
\caption{Digital Signature Generation}
\label{fig:digital-signature-building}
    \end{center}
\end{minipage}
\hfill
\begin{minipage}[t]{.45\textwidth}
    \begin{center} 
\epsfig{file=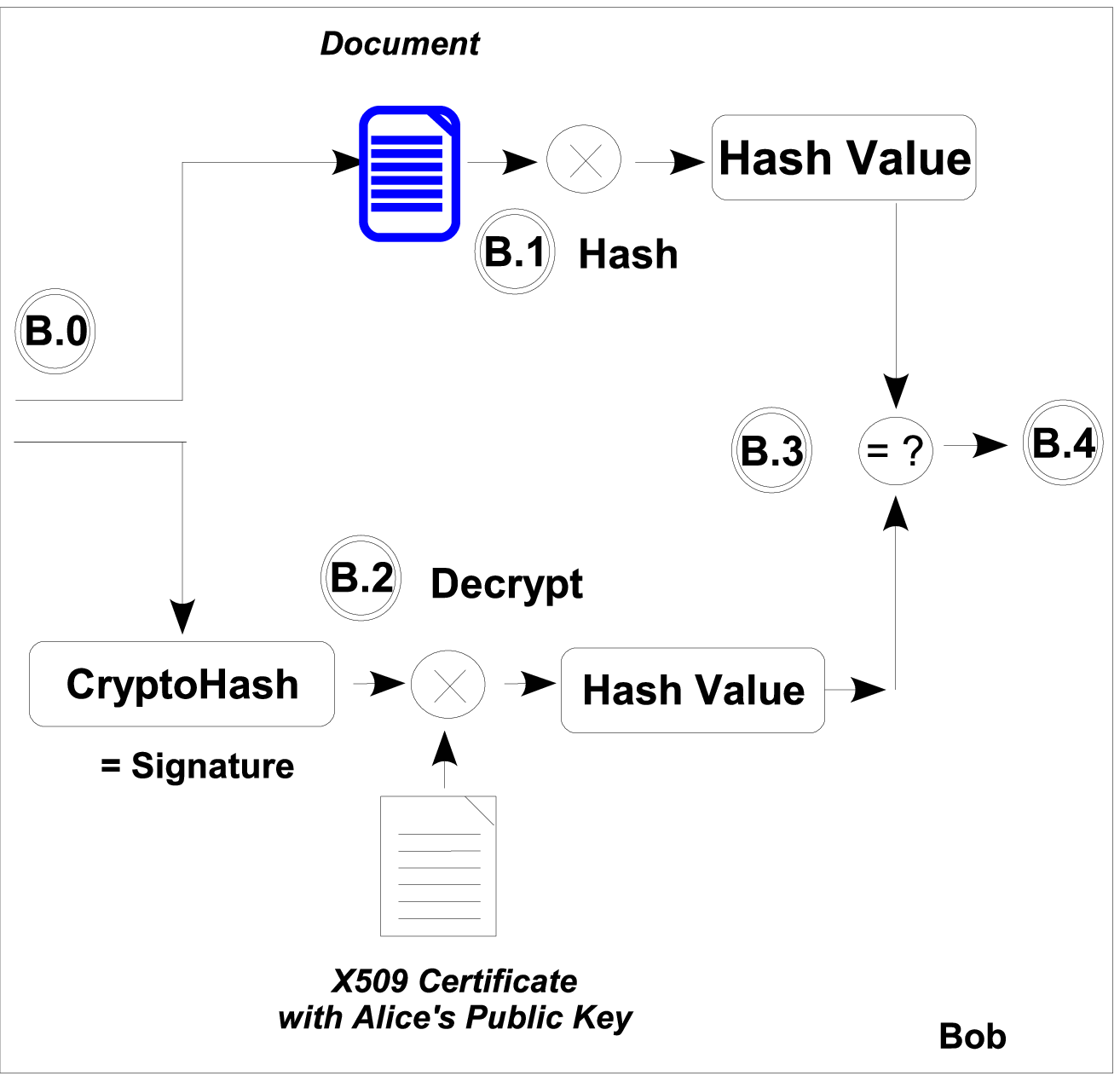, width=170pt}
\caption{Digital Signature Validation}
\label{fig:digital-signature-checking}
    \end{center}
\end{minipage}
\hfill  
\end{figure}

The figure \ref{fig:digital-signature-building} and \ref{fig:digital-signature-checking} show respectively the process of digital signature generation and the process of digital signature validation.

The process of digital signature generation takes as input the document to be signed and the private key of the signer. It produces as output a digital signature, or Cryptohash, which accompanies the signed document so as to prove its integrity. The first step of digital signature generation is to apply a hash function to the document, so as to obtain a fixed length data file that uniquely matches the original document (A.1). The resulting hash value is encrypted with the private key of the signer, so as to guarantee that nobody else could have produced this signature, and thus to prove that no malicious entity have provided or modified the document (A.2). The document can then be publicly published along with its digital signature (A.3).

When a user wants to exploit a document that is publicly available, or that has been transfered over an insecure communication channel such as the Internet, it can then verify that this document has been issued by the pretended issuer, and has not been tampered with during transfer. The process of digital signature validation takes as input the published document, the digital signature and the public key of the pretended signer (B.0, B.2). The validation is made of two parallel processes. On the first hand, the document is hashed with the same hash function that has been used for the signature generation (B.1). This step provides the hash value of the available document. On the other hand, the digital signature (=the Cryptohash) is decrypted with the public key of the signer (B.2). The hash value of the original document is retrieved by this way. This step guarantees that no other person tries to impersonate the real signer. Once the hash of the original document and the hash of the available document are available, it is sufficient to compare them. If they match, the available document is the one that has been signed. Otherwise, it means that the available document is not the one that has been signed.

The unvalidity of the process of digital signature validation can have several causes. The more obvious one is of course the modification of the document, or the substitution by another one. But the modification of the digital signature itself has the same result. If someone has a valid document without the original digital signature, it can not check the validity of the document. Another cause of validation error is the lack of knowledge about the signer. When the receiver does not have a copy of the public key of the signer that he knows to be valid, it cannot validate the signature. Actually, anybody can sign the document and provide a valid signature for it. If you do not trust the signer, the digital signature can not provide the proof of integrity of a document.

  \subsubsection{Cryptographic Message Syntax}

Another constraint exists in the verification of a document. The receiver of the document must have all necessary data for executing this validation, that is to say the document, the digital signature and the public key certificate of the signer. However, it can be necessary to trust document issuers that are not known beforehand, that is to say for which the public key certificate is not previously stored by the receiver. Therefore, this public key certificate must be provided along with the signed document. Process of trusting previously unknown signers implies signature delegation, which is presented in subsection \ref{authentication}. 

This data availability constraint means that several files must be transfered along with the signed document for verification. It is therefore necessary to bind them together, so as to prevent complex and slow document exchange protocols between the signer and the receiver. A solution for providing the document, the digital signature and the public key certificate of the signer is to integrate them in a CMS (Cryptographic Message Syntax) document \cite{rfc3369}\footnote{CMS is a follow up to PKCS7 message format, defined by RSA Laboratories \cite{pkcs7}}, and to publish not directly the signed document, but its associated CMS file. This CMS file can contain or not the signed document, depending on the context of publication.

\begin{figure}[htb]
\centering
\epsfig{file=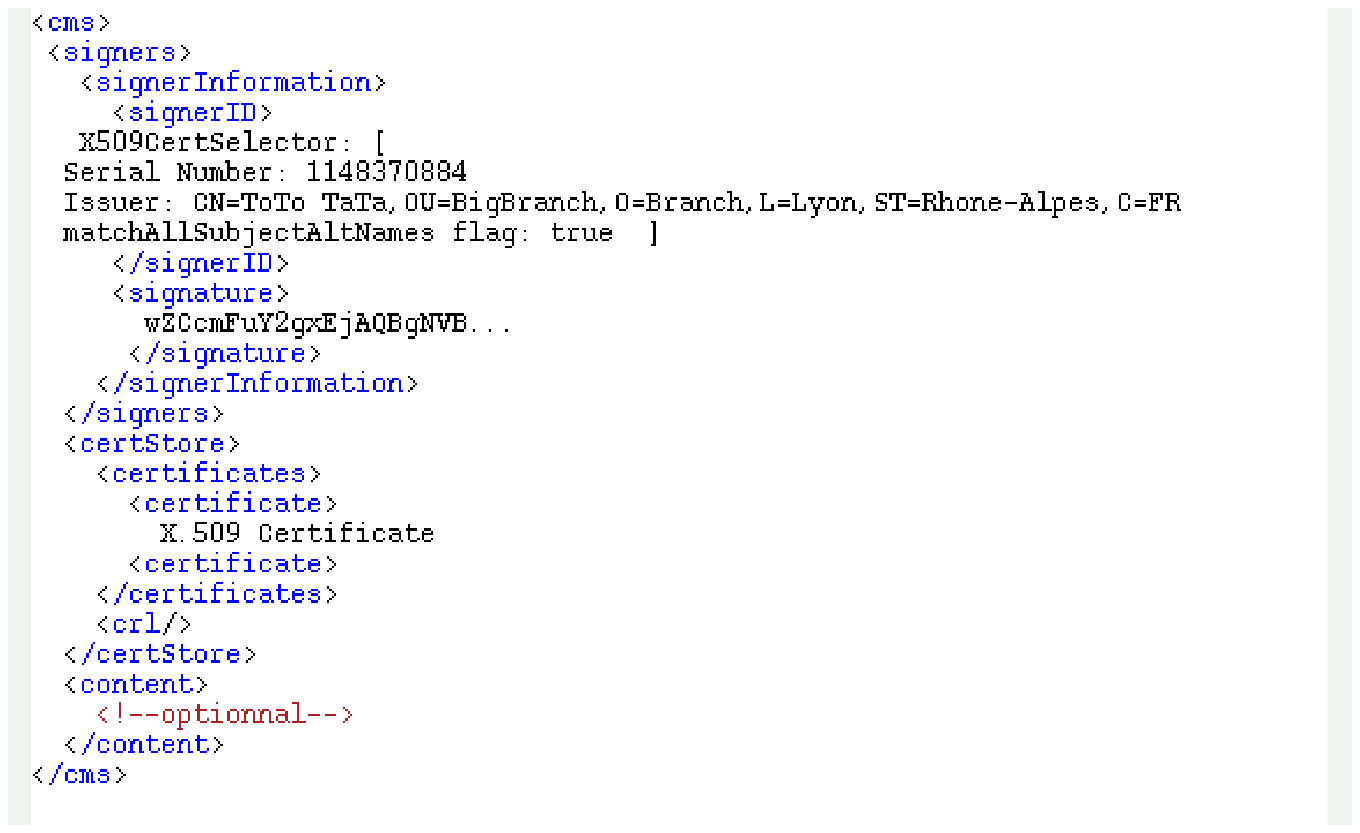, width=300pt}
\caption{The content of a Cryptographic Message Standard (CMS) compliant File.}
\label{fig:cms}
\end{figure}

Figure \ref{fig:cms} shows an example of a Cryptographic Message Syntax (CMS) compliant File in a human-readable XML format.

%key management techniques : ktri, kari, and kekri, pwri, ori

For encapsulating Digital Signatures, the CMS type `signed-data' is used. It contains data necessary for validating the Digital Signature. Encapsulated data in this kind of CMS file is organized into two categories: Signers data, and Certificates data. Signers data contains  the ID of one or several signer(s) of the document, as well as the Digital Signature of the document for each signer. Certificates data comprises X.509 Public Key Certificates for the signers, and potentially a Certificate Revocation List. Moreover, the signed-data file can also contain the signed document itself.

CMS Files use the ASN.1 syntax and are encoded in DER (Distinguished Encoding Rules) format \cite{kalisti93format}. This makes it possible to integrate binary content such as Digital Signatures together with the name of the signer and the properties of the certificates.

 \subsubsection{Asymmetric Encryption and Hashing Algorithms}

The process of digital signature generation and validation makes use of two different kinds of algorithms: one hash function, and one encryption algorithm. Numerous cryptographic algorithms are available that provide one or the other functionality. Although current recommendations for digital signature strongly restrict the choice between the RSA/MD5 pair and the DSA/SHA-1 pair, the choice of algorithm remains open. Actually, the required security level, the memory and performance constraints of the system that makes use of digital signature can strongly impact the choice.

Commonly used Hash algorithms are MD5 (Message-Digest algorithm 5) \cite{rfc1321}, SHA-1 (Secure Hash Algorithm) \cite{fips180}, as well as SHA-224, SHA1-256, Tiger \cite{anderson96tiger} and Whirlpool \cite{iso10118-3:2004}\footnote{Only SHA-1 or better algoritms should be used for signing documents, since it is possible to build false Certificates using MD5 in a couple of hours \cite{wang05md5}. Theoretical attacks also exist on the SHA-1 hash function, but are currently not exploitable in practice.}. Table \ref{tab:hash} shows the characteristics of these algorithms. The output length is the length of the resulting hash value. When several variants exist for a given algorithm, several output lengths are given. The security level indicates whether the algorithm is considered as secure. The availability indicates when a given algorithm is available in tools for signing archives (XXX), regularly considered in specifications (XX) or available in APIs but not integrated in existing tools (X).

\begin{table}[htb]
\begin{center}
\begin{tabular}{|l|c|c|c|}
\hline
\textbf{Hash Algorithm} & \textbf{Output length} & \textbf{Security} & \textbf{Availability}\\
& \textbf{(bytes)} & \textbf{Level} & \\
\hline
MD5 & 128 & Broken & XX\\
SHA-1 & 160 & Theoretically & XXX \\
& & Broken & \\
SHA-224 & 224 & High & X\\
SHA256 & 256 & High & X\\
Tiger & 128/160/192 & High & X\\
Whirlpool & 512 & High & X\\
\hline
\end{tabular}
\end{center}
\caption{Main Hash algorithms}
\label{tab:hash}
\end{table}

Commonly used encryption algorithms are RSA \cite{pkcs1}, DSA (Digital Signature Algorithm) \cite{fips186}, and ECC (Elliptic Curve Cryptography) \cite{rfc3278}. Table \ref{tab:encryption} shows the characteristics of these algorithms. The typical key sizes gives the key sizes commonly used for ensuring secure communications. For each key, different lengths can be used depending on the necessary security level. Several values are then given. The security level indicates whether the algorithm can be considered as secure. The availability indicates when a given algorithm is available in tools for signing archives (XXX), regularly considered in specifications (XX) or available in APIs but not integrated in existing tools (X). When available, the typical associated hash algorithm gives the hash function that is commonly used with the encryption algorithm.

\begin{table}[htb]
\begin{center}
\begin{tabular}{|l|c|c|c|c|}
\hline
\textbf{Encryption Algorithm} & \textbf{Typical} & \textbf{Security} & \textbf{Availability} & \textbf{Typical associated}\\
& \textbf{Key Sizes} & \textbf{Level} & & \textbf{Hash algorithm} \\
\hline
RSA & 1024/2048 & High & XX & MD5 \\
DSA & 1024/2048 & High & XXX & SHA-1 \\
ECC & 160/192  & High & X &  \\
\hline
\end{tabular}
\end{center}
\caption{Main Encryption algorithms}
\label{tab:encryption}
\end{table}

The choice of pairing a hash function with an encryption algorithm is quite open, although one restriction exists: the length of the data generated through the hash function must be at least as long as the encryption key. For instance, SHA-1 generates a digest of 160 bytes (=1280 bits). The longest key can then be 1024 bits. For a more powerful signature, it is necessary to use a hash function with longer output, for instance SHA-256 (256 bytes = 2048 bits). A key of 2048 bits can then be used.

For non strategic systems, the current trend is to use the DSA/SHA-1 algorithm pair. It is considered as sufficiently secure, and has the non-negligible advantage of compatibility with existing signature tools.

\bigskip

The digital signature of a document enables to guarantee its integrity, that is to say that a given document is identical to the original one. In particular, this aims at preventing document modification or substitution. However, the guarantee of integrity requires that the receiver trusts the signer of the document. Consequently, it is necessary to authenticate the signer. Otherwise, anybody can publish a file and provide a valid signature. Two cases of authentication exist: either the signer is already known to the receiver, or it is not, and a trusted third party is then required to guarantee the identity of all potential signers. Following section will detail the authentication process.

 \subsection{Authentication}
\label{authentication}

% \subsubsection{Definition}

\emph{Authentication} is the formal identification of the emitter of a message. %In the context of Digital Signature, it is the formal identification of the signer of a document. 
It consists in the verification of the validity of the identity of this emitter.

In the context of Digital Signature, the emitter of the message is in fact the signer of the document. Its identity is carried along with the signed document as a public key certificate compliant with the X.509 format. These data are encapsulated together in a CMS file. The authentication process implies to compare this certificate bound to the document and the certificates that are considered as trusted by the entity that performs the authentication. These trusted certificates are stored in a database called Certificate Store.

  \subsubsection{Certificate Validation}

Two scenarios of authentication of a public key certificate exist. Either the Subject, that is to say the signer of the public key, is known to the entity that performs the authentication, or it is not. In the second case, the authentication can be performed if the Issuer of the public key certificate (or one issuer of the issuer's certificate) is known.

\begin{figure}[htb]
\centering
\epsfig{file=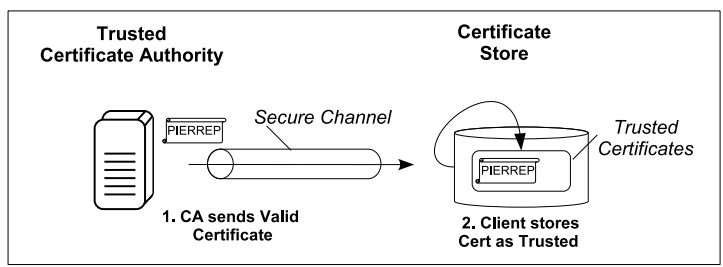, width=270pt}
\caption{The Certification Distribution required for Certificate Checking}
\label{fig:initializeCerts}
\end{figure}

The requirement for both authentication scenarios is that the set of certificates that are considered as trusted are transfered in a secure way to the entity that performs the authentication before the authentication occurs. The figure \ref{fig:initializeCerts} shows this process.

The distribution of trusted certificates is done in an initialization phase. A trusted Certification Authority delivers the set of certificates that the authenticating part can trust over a secure channel (or possibly offline). These certificates are stored by the authenticating part in its Certificate Store, and marked as trusted. This means that when the client latter handles a certificate that is available in the Certificate Store and is marked as trusted, it will be able to assert that this certificate is a valid one. In other cases, it will not be able to verify whether a given certificate which Subject is for instance SFRENOT is a valid one, or a fake one built by someone who pretends to be SFRENOT but who is not.

\begin{figure}[htb]
\centering
\epsfig{file=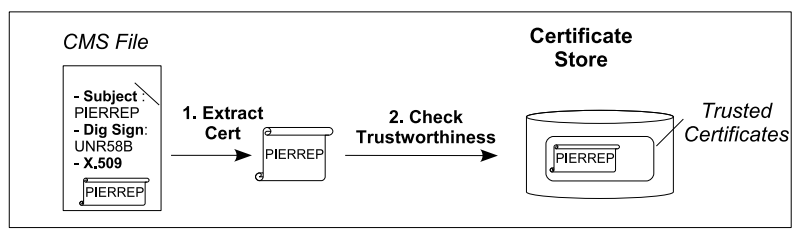, width=300pt}
\caption{Certificate Validation (Case 1): the Certificate is known to the checker}
\label{fig:knownCertVal}
\end{figure}

The first mechanism (Case 1) is shown on figure \ref{fig:knownCertVal}.

In the Case 1, the authenticating part first need to retrieve the public key certificate. In particular, in the case of Digital Signature transmitted through a CMS file, it extracts the certificate from it (1). Then it can check whether this certificate is already known and marked as trusted (2). In our example, the signer is called PIERREP. The Subject of the certificate is then PIERREP. It is possible to assert its validity because the same certificate for PIERREP (with same Issuer and certificate signature) is marked as trusted in the Certificate Store.

\begin{figure}[htb]
\centering
\epsfig{file=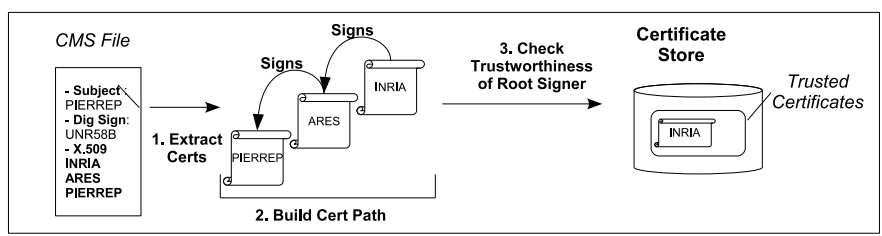, width=330pt}
\caption{Certificate Validation (Case 2): the Certificate is unknown to the checker}
\label{fig:unknownCertVal}
\end{figure}

The second mechanism (Case 2) is shown on figure \ref{fig:unknownCertVal}.

The second mechanism is made possible by the structure of public key certificates. They are either issued by a third party, or self signed. In any case, they contain two identities that are relevant for authentication. The first one is the identity of the Subject, that is to say the owner of the certificate. The second one is the identity of the Issuer, that is to say the entity that has emitted the certificate.

The authenticating part first need to retrieve the various public key certificates that are required to perform the authentication (1). Then it builds the certificate path, by linking the certificate of each Subject with the one of its Issuer (2). The validity of the certificate path is asserted if the root certificate that is to say the first of the certificate hierarchy exists in the Certificate Store and is marked as trusted (3). The certificate path is only valid if all certificates used for signing other ones have the right to do it, which is indicated in the certificate itself. In our second example, the signer PIERREP is unknown to the authenticating part. It is provided with the certificate for ARES, who has been used to issue it, and the certificate for INRIA, that has been used to sign the ARES one. The validity of this certificate chain is asserted because both INRIA and ARES have the right to issue certificates, and because the authenticating part knows the certificate of INRIA.

  \subsubsection{X.509 Certificate}

The X.509 public key Certificate is a data structure that encapsulates a public key and associated data necessary for identifying a given subject \cite{rfc2459}. It can be published in a CMS file, or as stand-alone data.

 The X.509 Certificate is digitally signed by the issuer of the public/private key pair, which thereby claims that the subject of the certificate is the owner of the associated private key. It also claims that he acknowledges that the subject's name (called Distinguished Name) is correct. Every user which trusts the issuer of the certificate will then be able to assert the identity of the emitter of a message that can be decrypted (or whose digital signature can be verified) with the public key contained in this certificate. This is the authentication. This process implies of course that the private key has not been corrupted.

\begin{figure}[htb]
\centering
\epsfig{file=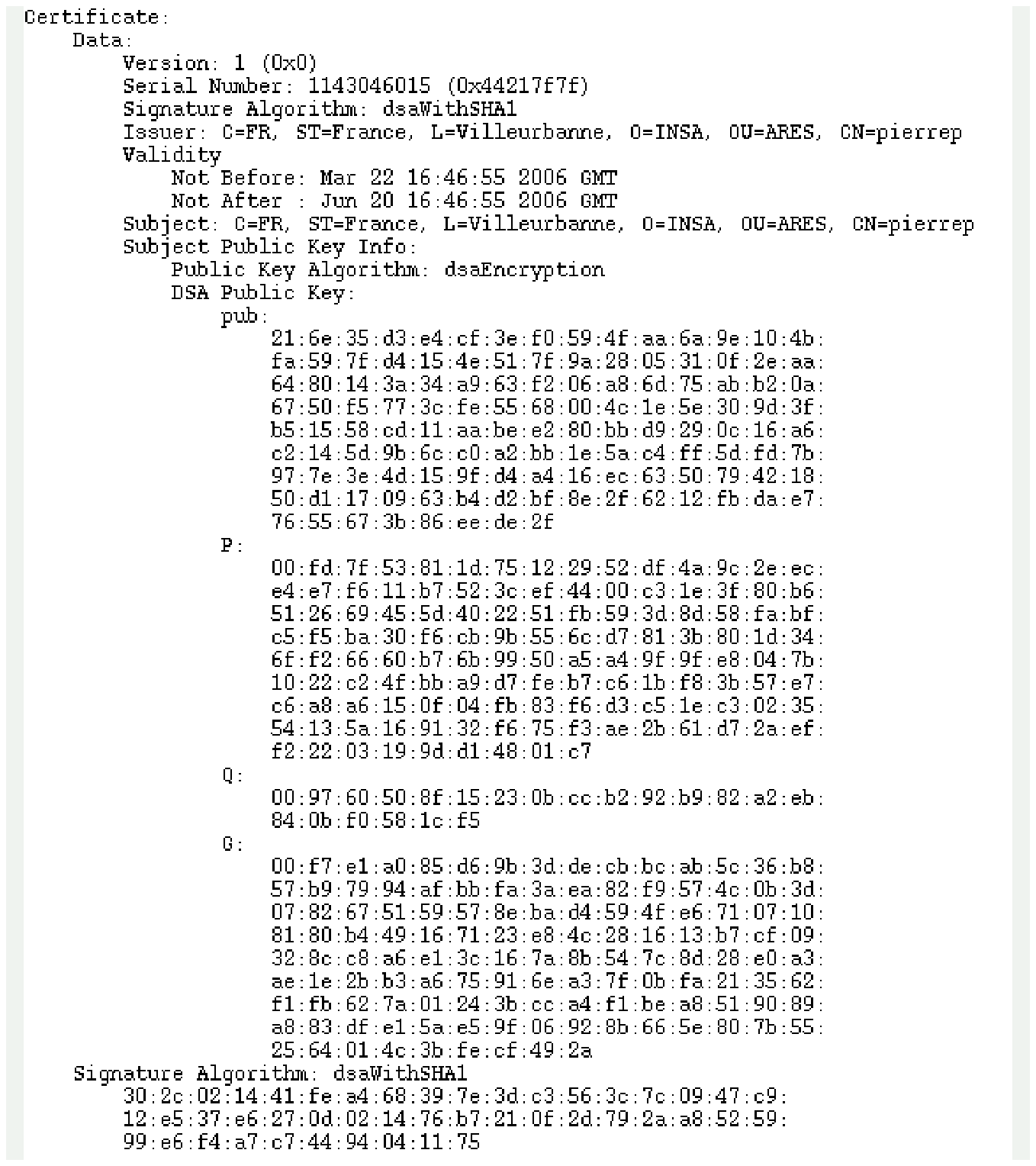, width=250pt}
\caption{Content of a X.509 Certificate}
\label{fig:X509}
\end{figure}

 Figure \ref{fig:X509} shows the content of a X.509 certificate.

Fields of the certificate are the public key, the name of the Subject, the name of the Issuer of the Certificate, the validity period, the URL of the revocation server, and the Digital Signature. A additional field allows to check the validity of the certificates, and is not represented in this human-readable representation: the digital signature of the certificate by its issuer.

The name of subjects are defined by Distinguished Names (DN), which originate in LDAP specifications \cite{rfc2253}. Distinguished Names are a composition of following fields: CN=`Common Name', OU=`Organization Unit', O=`Organization', L=`Location'(city), S= `State', C=`Country code'. Definition of Distinguished Names states that the fields follow a hierarchical organization, and thus that their order imports. For instance, a DN \{C=France, O=INRIA\} is different of a DN  \{O=INRIA, C=France\}. However, in several tools for manipulating certificates, such as Sun Keytool, the order of the fields is fixed, preventing such ambiguities.

  \subsubsection{Certificate Store}

A Certificate Store is a database that contains Public Key Certificates that the owner of the Store knows. Certificates can be identified as trusted and untrusted. It usually also includes one or several private keys. It is then called a Keystore\footnote{http://java.sun.com/j2se/1.5.0/docs/tooldocs/solaris/keytool.html}.

The Certificate Store is protected by a password. When private keys exists, each one is protected by its own password. Consequently, a Certificate Store (or a Keystore) can be shared among several subjects, if the certificate management is shared.

\begin{figure}[htb]
\centering
\epsfig{file=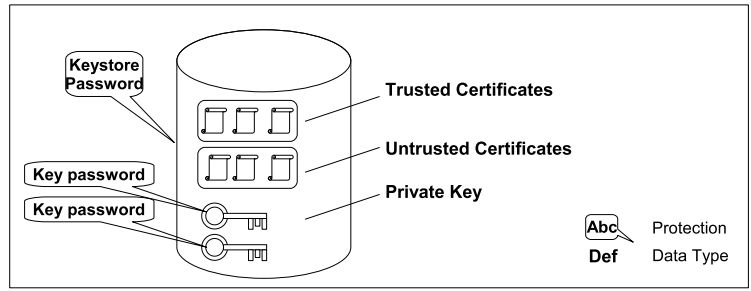, width=280pt}
\caption{Content of a Certificate Store}
\label{fig:certstore}
\end{figure}

Figure \ref{fig:certstore} shows the content of a Certificate Store: trusted certificates, untrusted certificates, private keys.

\bigskip

The conjunct use of a digital signature and a certificate store allows to guarantee both the integrity of a document and to authenticate its signer. Both properties must apply together. However, this mechanism, if it guarantees that a document has not been modified after its publication, does not protect its content from malicious eavesdroppers. The third security property, confidentiality, can be used to achieve such a protection.

 \subsection{Confidentiality}

  \subsubsection{Definition}

\emph{Confidentiality} is the absence of unauthorized disclosure of information \cite{avizienis00dependability}. In the context of component deployment, this means that the content of the component - code or other resources - is not available to users that are not explicitly authorized to manipulate them.

  \subsubsection{How to achieve Confidentiality ?}

Confidentiality with asymmetric cryptography is achieved by encrypting the document with the Public Key of the receiver (1). Thus, the owner of the matching private key is the only one that can decrypt the document (2), and gain access to its content.

\begin{figure}[htb]
\centering
\epsfig{file=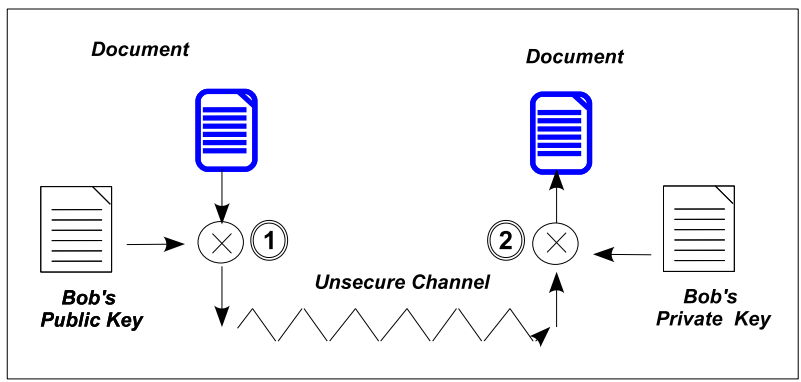, width=300pt}
\caption{Encryption of a document for confidentiality}
\label{fig:confidentiality}
\end{figure}

Figure \ref{fig:confidentiality} shows the process of encryption and decryption of a document for ensuring confidentiality.

Because the document is not encrypted with the private key of the emitter,  this process does not provide means of performing authentication. In practice, encryption for confidentiality is used together with encryption for authentication. Therefore, the receiver of the document, after having decrypted it with its own private key, can make the decryption with the public key of the emitter and thus control its identity.

\subsection{Conclusions}

Security requirements for computer systems are \emph{integrity}, \emph{authentication} and \emph{confidentiality}. The first property aims at verifying that the content of data have not been tampered with. The second property aims at identifying the emitter of the data. Both steps can not be considered independently: if the emitter of a message is not authenticated, any malicious user can publish data with correct integrity validation mechanism. Similarly, if a message is authenticated but its integrity is not guaranteed, there is no way not know whether any single byte of it has really been emitted by the authenticated emitter.

The third property is confidentiality. It aims at protecting data from undue read access. In this report, we will not consider it further since we consider that access to component resources does not make it possible for a malicious user to gain access to the component platform.

%% file: secureDeployment.tex
\section{Secure Deployment}
\label{secureDeployment}

The important increase of quantity and diversity in component systems makes it necessary to protect them against malicious persons and systems. Components can be modified during deployment, or simply be published by untrusted issuers. It is therefore necessary to guarantee the integrity of the components and the authentication of their issuer. Confidentiality will not be considered further.

Security mechanisms must not imply modifications in the deployment process. Users (and platforms) must continue to use their component platforms and to update it without modification. Consequently, the signed components must be delivered as a single archive which includes both the original resources and the data necessary for verifying integrity and authentication. Otherwise, the security mechanisms are not transparent, they will be rejected by the users, and will not help improve the security of the systems. 

In the case of OSGi$^{tm}$ platforms, the solution consists in including the digital signature in the component (also named bundle) itself. Consequently, it is not possible to sign the whole component. A list of resources present in the component and of their respective hash value is built. This list is the one that is signed, and included in the meta-data of the component along with the signature.

This section first presents an overview of the entities that intervene in the secure deployment of bundles, of the requirements and of the process for signing and verifying signed components. Next, it details the structure of a signed bundle as defined for OSGi$^{tm}$ bundles. Lastly, the process of signature generation and signature validation will be presented.

 \subsection{Overview}
\label{secureDeploymentOverview}

This subsection provides an overview of the entities and data that intervene in secure deployment of bundles in an OSGi$^{tm}$ platform. It shows how the digital signature can be exploited in the context of bundle deployment, so as to guarantee the integrity of a bundle and the authentication of its signer.

 First, the principles of bundle deployment are presented. Then, the requirements for supporting a digital signature based security mechanism are presented, as well as the overall process of bundle signing and validating.

  \subsubsection{Bundle Deployment}

The deployment of a bundle is the part of its life cycle that spans between the end of its development and the moment it is ready to provide services on an OSGi$^{tm}$ platform. It contains several steps: publication of the bundle, discovery, dependency resolution, download, installation, configuration. Update phase must also be taken into account \cite{hall99cooperative}. 

Two main types of deployment can be identified. The first kind is platform-initiated deployment, which can be seen as 'pull deployment', where the signal that triggers the deployment originates from the component platform itself. The second kind of deployment occurs when it is initiated through a remote signal. This occurs for instance in the case of console-based remote management of a component platform \cite{royon06gestion}.

The entities that intervene in the deployment process are the following:
\begin{description}
 \item [The Bundle Issuer,] it is the person or system that makes the bundle available for the end users. It can be a software developer or a software vendor.
 \item [The Bundle Repository,] it is the server that publishes the bundles, that is to say that provide a remotely accessible service so that the end users can find and download the bundles.
 \item [The Execution platform,] it is the component platform that executes the bundles. It must support bundle deployment, and often initiates it. In this report, it is also simply called the client.
\end{description}

The deployment process is initiated by a deployment trigger. This trigger is either a local shell (pull deployment) or a remote console (push deployment).

Securing the deployment process implies that the execution platform only deploys and installs bundles that come from trusted issuers. As far as bundles are signed, they can safely be published on insecure repositories. The deployment trigger, on its side, need to have a secure communication channel to the platform, so as to prevent deployment of bundles by untrusted parties. Moreover, it must be protected from undue use. The protection of deployment trigger relates to system management more than to deployment, so it will not be studied further here. We assume that only valid users have access to the execution platform.

  \subsubsection{Requirements for Authentication}

Subsection \ref{authentication} has shown that the condition for exploiting digital signature as a mean of proving both integrity of a document and authentication of its signer is that the entity that checks the signature (we will call it the client) knows either the public key of the signer itself, or the public key of the certificate issuer that has provided the signer's key. Through signature delegation, a complete hierarchy of certificate issuers can exist between the certificate issuer the client knows and the signer itself.

In any case, the signer must have a private key that the client can trust, and the client must have a public key that he knows to be trustworthy. Typically, both keys are provided by a common certification authority through a secure communication channel. 

\begin{figure}[htb]
\centering
\epsfig{file=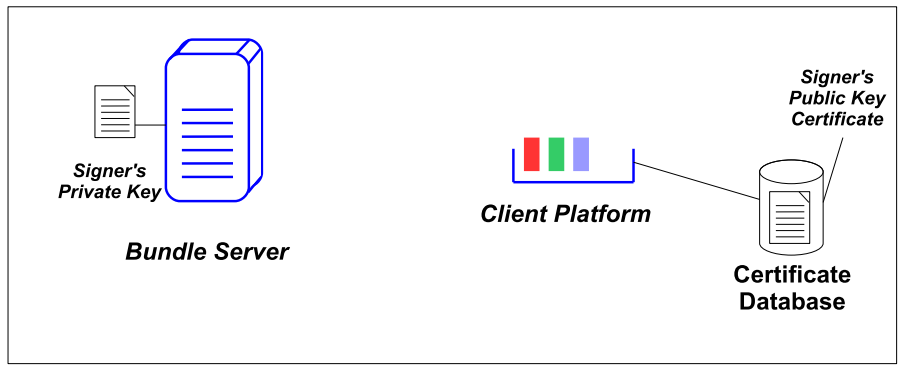, width=350pt}
\caption{The requirements for the authentication process.}
\label{fig:authentication-req}
\end{figure}

Figure \ref{fig:authentication-req} shows the requirements for the authentication process: the public key of the authenticated part must be known to the entity which wishes to perform the authentication.

Once the requirement of key availability is fulfilled, the secure deployment can occur.

Next section will present the internal structure of a signed bundle, and the way the digital signature is used to sign not only one document, but also all available resources of the bundle.

 \subsection{Structure of a Signed Bundle}
 \label{bundleStructure}

Because of the particular constraints on the signature of a bundle, it is necessary to store it and all related resources in the bundle itself. Moreover, it is mandatory that multiple signers can sign the same bundle.

\begin{center}
\begin{figure}[htb]
\epsfig{file=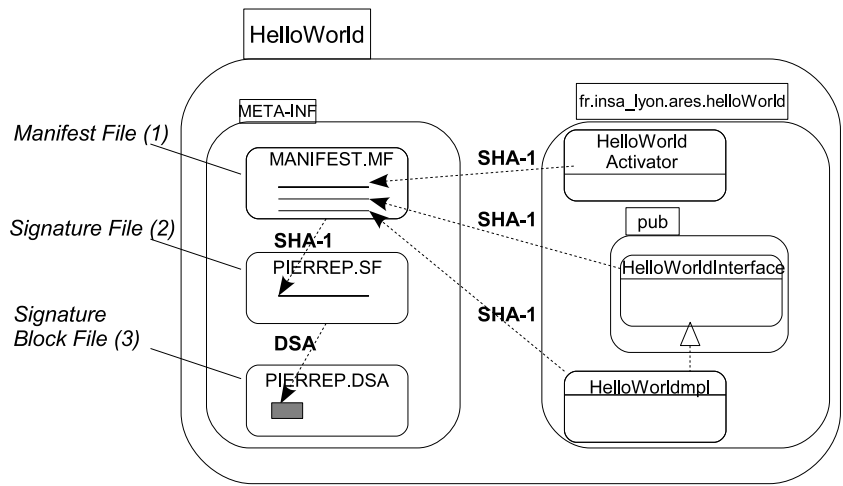, width=320pt}
\caption{Example of a signed HelloWorld bundle, signed by PIERREP.}
\label{fig:bundle-signed-example}
\end{figure}
\end{center}

The structure of a signed bundle is shown on Figure \ref{fig:bundle-signed-example}.

\begin{center}
\fbox{
\begin{minipage}[c]{10.5cm}
\centering
\emph{Signature File and Digital Signature}
\begin{itemize}
 \item The \textbf{Signature File} of a signed bundle is a meta-data file that the contain the Signature of the \textbf{Manifest file}, that is to say its hash alue. It guarantees the integrity of the \textbf{Manifest file}.
 \item The Digital Signature of a file is a byte array that contains the signature of a given file by a given person, that is to say the encryption of the hash value of the signed file. In a signed bundle, the Digital Signature of the so-called \textbf{Signature File} is stored in the \textbf{Signature Block File}. It guarantees the integrity of the \textbf{Signature File} and the identity of the signer
\end{itemize}
\end{minipage}
}
\end{center}

A first solution for archive signing (bundles are specific jar archives) is given by the Jar Archive specifications \cite{jarSpec}. However, this gives the possibility to sign only a subset of an archive. This implies that modifications are possible on the archive even after its signature, which is a potential security leak. Therefore, OSGi$^{tm}$ specifications restrict the signature by imposing that all resources in an archive are signed by the signer(s). In the contrary case, the signature is not valid. Embedded archives must be signed on the same way. OSGi Signature Files only need to contain the hash value of the Manifest, hash value of the other resources are not required \cite{osgi05core}.

 The \textbf{Manifest file} of the archive (1) contains the hash value of each resource in the archive. To support several signers, the digital signature is applied not directly on the \textbf{Manifest file}, but on a so-called \textbf{`Signature File'} (2), which is specific to each signer. A hash value of the \textbf{Manifest file} must be included. The digital signature of this \textbf{Signature File} is stored along with data that are necessary for its validation in a CMS file of type `signed-data' which is named \textbf{`Signature Block File'} (3).

This structure of a signed bundle will be enlightened by a simple example of the HelloWorld bundle, whose signer is named PIERREP. This bundle contains three classes: HelloWorldActivator (the activator, or starter, of the bundle), HelloWorldInterface (the definition of the HelloWorldInterface service that is provided by the bundle), and HelloWorldImpl (the implementation of the above mentioned service).

The meta-data of the bundle are the following. First, the \textbf{Manifest file}, \textbf{MANIFEST.MF}, which contains meta-data specific to OSGi bundles, as well as the hash value of all resources. Secondly, the \textbf{Signature File}, \textbf{PIERRE.SF}, contains the hash value of the \textbf{Manifest file}. Thirdly, the \textbf{Signature Block File}, \textbf{PIERRE.DSA}, is a CMS file that contains the digital signature of the \textbf{Signature File}, and the public key certificate of the signer. They must be stored in this order (and before all other resources) in the bundle archive.

A overview of the three meta-data files is shown is table \ref{tab:metadata}. Specific characteristics of each if the meta-data files used for bundle signature are presented in subsequent paragraphs.

\begin{table}[htb]
\begin{center}
\begin{tabular}{|c|c|c|}
\hline
\textbf{File denomination} & \textbf{Example} & \textbf{Content} \\
\hline
Manifest File & MANIFEST.MF & Hash value for each resource in archive\\
Signature File & PIERREP.SF & Hash value of the Manifest File\\
Signature Block File & PIERREP.RSA & Digital Signature of Signature File\\
\hline
\end{tabular}
\end{center}
\caption{Meta-data involved in Bundle Signature}
\label{tab:metadata}
\end{table}

  \subsubsection{The Manifest File}

\begin{figure}[htb]
\centering
\epsfig{file=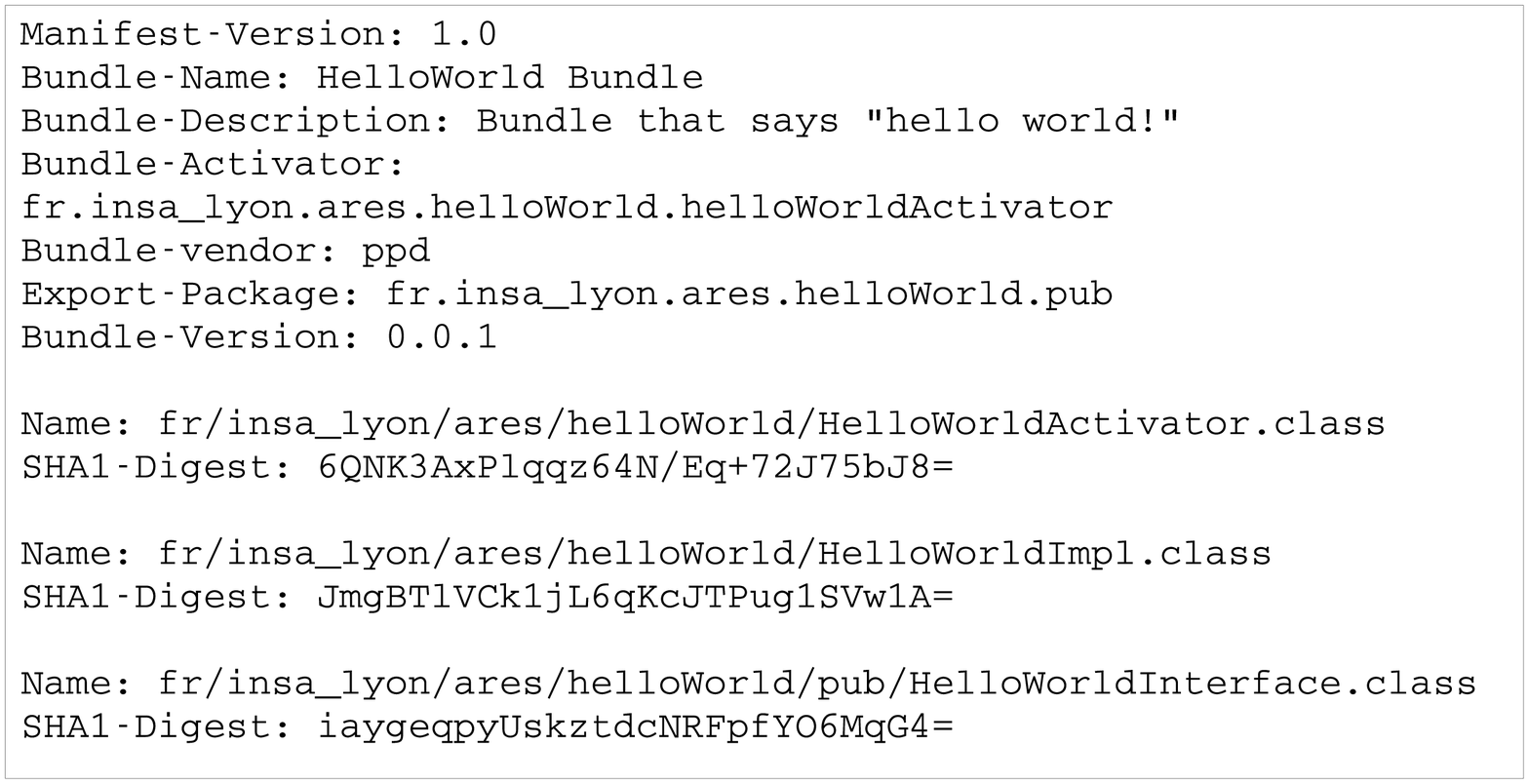, width=350pt}
\caption{The Manifest File for the HelloWorld Example Bundle.}
\label{fig:manifest-example}
\end{figure}

The Manifest file for the HelloWorld example bundle is shown in Figure \ref{fig:manifest-example}.

The Manifest file of an OSGi$^{tm}$ bundle contains the meta-data required for bundle deployment: the name of the bundle, its version, the packages it provides, the packages it depends on, its activator class for starting it. In a signed bundle, this meta-data is enriched by the hash value of each resource the bundle contains. A resource is identified by its full path inside the bundle. It is completed by a property that indicates its hash value. The property name depends on the hash function that is used. In our example, this hash function is SHA-1, and the matching property is `SHA1-Digest'. Note that a manifest file that contains resource entries that do not exist in the archive or that does not list all resources in the archive has probably suffered addition or removal of resources and is not valid.

Storing the hash values of the resources of the archive guarantees that none of this resources have been tampered with after the moment the bundle has been signed. Moreover, it guarantees that no resource have been added or removed.

  \subsubsection{The Signature File}

\begin{figure}[htb]
\centering
\epsfig{file=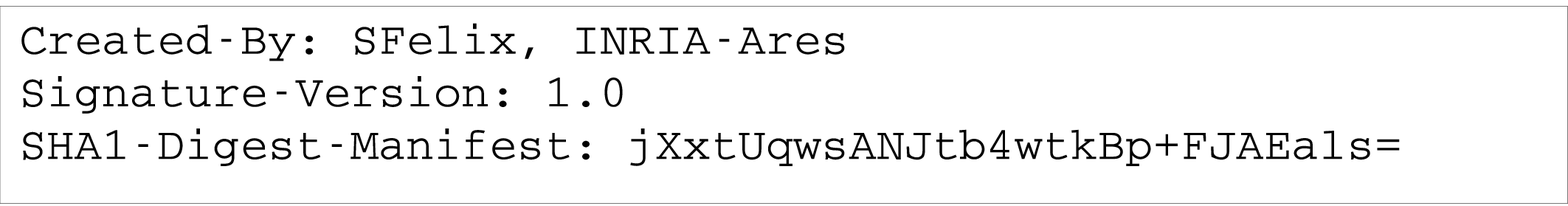, width=350pt}
\caption{The Signature File for the HelloWorld Example Bundle.}
\label{fig:signature-file-example}
\end{figure}

The Signature file for the HelloWorld example bundle is shown in Figure  \ref{fig:signature-file-example}.

Each signer of a bundle creates its own signature including both Signature File and Signature Block File. The name of those file is the capitalized name of the signer. The Signature File is identified by a `.SF' extension. For instance, in our example, the Signature File is named `PIERREP.SF'.

The Signature File of signed OSGi$^{tm}$ is simpler that the one of signed Jar archives. As far as it is not possible to sign a subset of the resource, no copy of the list of the name and hash value of the resources is required. Only the signature version and the hash value of the manifest is necessary. Hash function is usually the same that is used in the manifest for identifying resources. In our HelloWorld example, the hash value of the manifest is stored under the property name `SHA-1-Digest-Manifest'. 

Storing the hash value of the manifest file enables to guarantee that it has not been tampered with after the moment the bundle has been signed.

  \subsubsection{The Signature Block File}

The Signature Block File of a signed bundle contains the digital signature of the Signature File and all data that are necessary to check the validity of the signature. Its name is made of the capitalized name of the signer. The file extension is the named of the encryption algorithm used for the digital signature. It is therefore either `RSA' or `DSA'. In our HelloWorld example, the name is PIERREP.DSA.

The Signature Block File is a CMS compliant file (see subsection \ref{integrity}). It contains the public key certificate of the signer, and a SignerInfo data structure with the identifier of the signer and the digital signature itself. It can also contain a Certificate Revocation List.

The Signature Block contains a valid signature if the digital signature is a valid one for the Signature File, and has been created using the private key of the signer. Of course, the validation process must check whether the public key certificate can be trusted (see subsection \ref{authentication}). It guarantees that the Signature File has not been modified since the signature occurred, and that the signer is a trustworthy one.

The reader can refer in the Figure \ref{fig:cms} for an example of a Signature Block File.

\bigskip

Once the structure of a signed bundle and of its meta-data files have been presented, the algorithms used for signing the bundle, and for validating this signature, will be detailed. 

 \subsection{The Process of Signature and Validation}
\label{detailedProcess}

The process of signing bundle must create bundle meta-data that are compliant with presented specifications. Not only the meta-data content must be valid, but several other constraints must also be considered: the order of resources in the archive and the exhaustiveness of identified resources.

Of course, the process of validation of the bundle signature must check the same constraints.

  \subsubsection{Signature}

The main steps of bundle signature generation are the following. First, the public/private key pair must be available before signing. This is the initialization phase. Next, the manifest file, MANIFEST.MF, is generated. It contains the name of every resource in the bundle along with their hash value. Then, the Signature file is generated. It contains the hash value of the manifest file. The Signature Block File is generated, and contains the digital signature of the Signature File, and the public key certificate of the signer. Lastly, the whole archive is generated, the meta-data are sorted first, and then the other resources. 

\begin{figure}[htb]
\centering
\epsfig{file=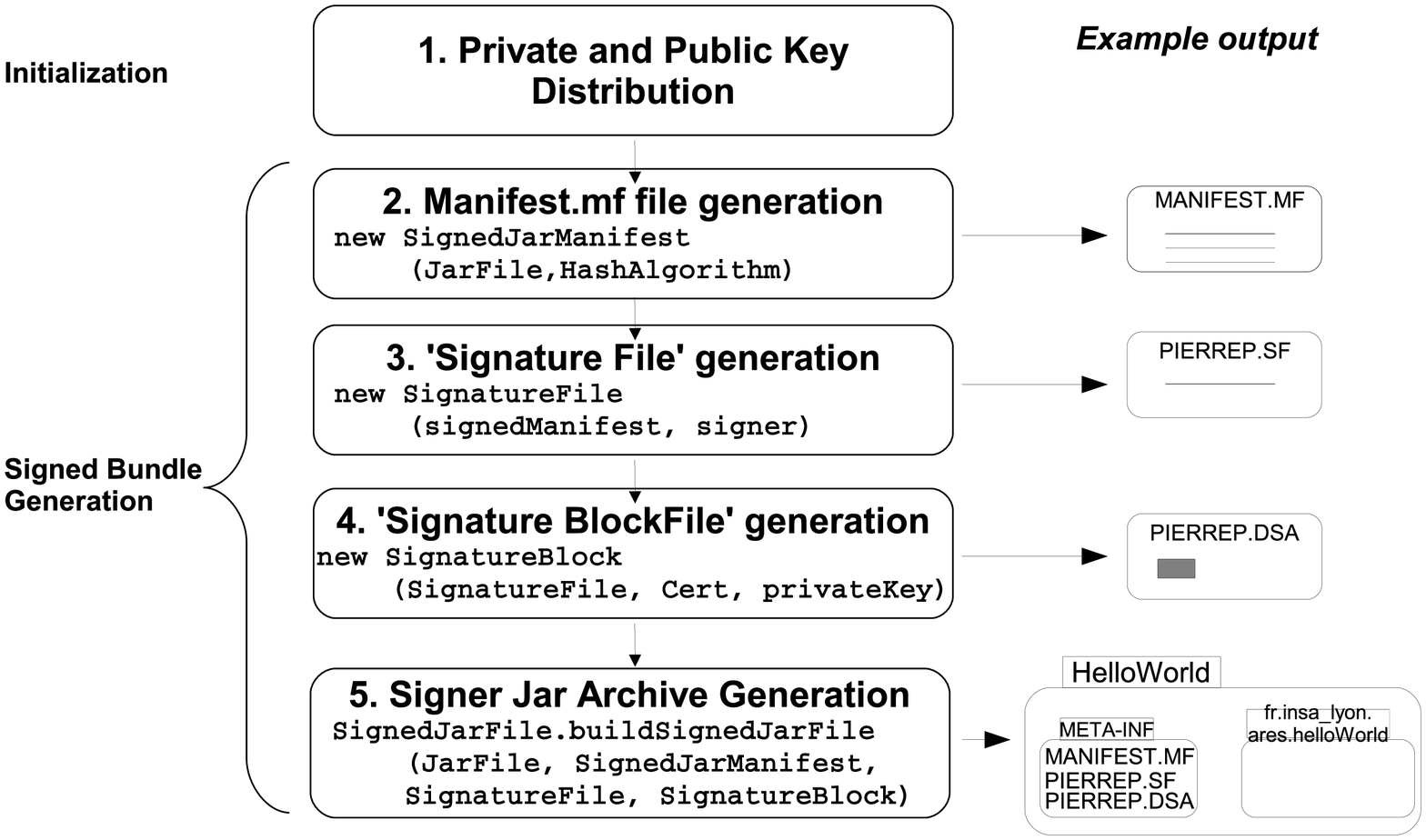, width=320pt}
\caption{The Algorithm for signing a Bundle.}
\label{fig:signature-algorithm}
\end{figure}

Figure \ref{fig:signature-algorithm} shows the algorithm for signing a bundle. You can refer to figure \ref{fig:seq-signing} from Annexe \ref{annexe:algos} for further details.

  \subsubsection{Validation}

The process of bundle signature validation is symmetric to the signature generation one. First, the entity that checks the signature needs to authenticate the signer, that is to say to check whether it knows its public key certificate or it is capable of establishing a Certificate Path between this public key certificate and a certificate he knows (see subsection \ref{authentication}). If the signer can not be authenticated, it is not worth trying to verify the signature, because anybody can build a valid signature.

The second step of the validation of bundle signature is the verification of the correct order of the resources in the archive. As already mentioned, the first files must be in this order the Manifest file, the Signature File and the Signature Block File. All other resources come afterwards.

The third step is the validation of the coherence of the meta-data files. The Signature Block File must contain a valid digital signature of the Signature File by the signer. The Signature File must contain the correct hash value for the manifest file. The Manifest file must contain the hash value for all resources of the archive, without exception, and without omission.

When these three steps are checked and valid, the signature of the bundle is valid. Should any of the criteria not be met, the bundle signature is not valid.

\begin{figure}[htb]
\centering
\epsfig{file=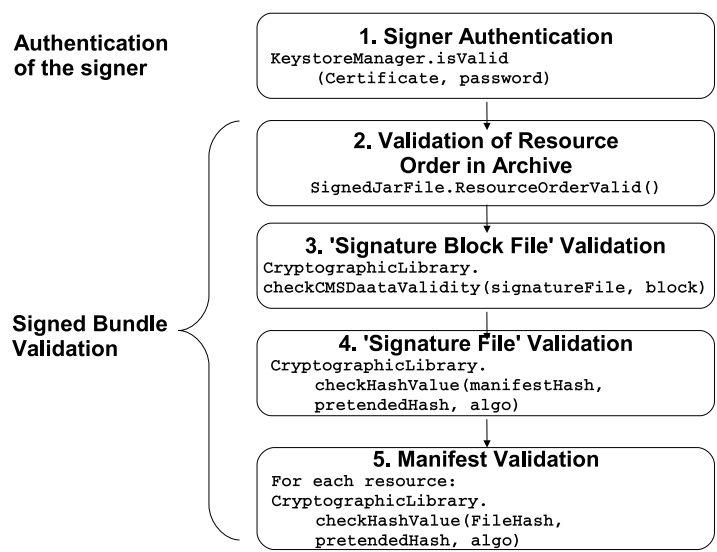, width=270pt}
\caption{The Algorithm for Validating a signed Bundle.}
\label{fig:validation-algorithm}
\end{figure}

Figure \ref{fig:validation-algorithm} shows the algorithm for validating a signed bundle.  You can refer to figure \ref{fig:seq-verification} from Annexe \ref{annexe:algos} for further details.

 \subsection{Conclusions}

Securing the deployment of components implies to protect the execution platform from malicious component publishers, and from potential modifications of the components after their publication. Such protection is achieved in the case of OSGi$^{tm}$ bundles by the signature of the bundles, which is based on digital signature and enables to store the signature itself and related data inside the bundle itself. The protection of bundles is done through two steps. First, the bundle is signed by a bundle publisher that is publicly known. Secondly, the bundle signature is validated just before being installed, so as to check that the signature is valid and that the bundle has not suffered modifications.

Such a process does not prevent malicious eavesdropper to steal the content of the component. This protection level would require confidentiality, and can not be achieved by simple integration of meta-data in the component. It requires encryption of the component, which makes necessary to have a dedicated key management facility. Moreover, it would break the compatibility of published components with unsecured platforms.

%% file: implementation.tex
\section{Implementation: SFelix}
\label{implementation}

SFelix\footnote{http://sfelix.gforge.inria.fr/} is the implementation of the bundle signature validation process of the OSGi$^{tm}$ specification, which is a part of OSGi$^{tm}$ Security Layer. It is provided together with the SFelix JarSigner tool, which enables to sign bundles and to publish them on a FTP server. It also provides the possibility to update the repository meta-data file for the OSGi$^{tm}$ Bundle Repository version 2 (OBR 2). SFelix is based on the Felix\footnote{http://incubator.apache.org/felix/} implementation of the OSGi$^{tm}$ platform. Felix is a project from the Apache Incubator, and is a follow-up of Oscar OSGi$^{tm}$ platform\footnote{http://oscar.objectweb.org/}.

To the extent of our knowledge, no other implementation of bundle signing and validation facility exists for the OSGi$^{tm}$ platform. SFelix is then the first project to provide it, at least in the Felix Project. Moreover, no Java implementation of a jar archive signer seems to be available as open source project. A tutorial exist on the OnJava web site, but uses Sun libraries that have been removed from the Java Virtual Machine distribution\footnote{http://www.onjava.com/pub/a/onjava/2001/04/12/signing\_jar.html}.

It has thus been necessary to implement the whole bundle signature and validation process in SFelix. An implementation of the algorithms for signature and validation have been developed (see subsection \ref{detailedProcess}).

 \subsection{Overview}

We first present an overview of the principles of the SFelix secure component deployment application. It is made up of the platform and of the JarSigner tool. The precise role of the application is detailed, then the structure of the program and its public API are explained.

  \subsubsection{Role of SFelix platform and SFelix JarSigner}

SFelix JarSigner and SFelix cover the whole deployment process of components. 

SFelix JarSigner covers the issuer side of the bundle deployment process. It allows a bundle issuer to sign the bundle, and to publish them on a public repository. Currently, only the FTP protocol is supported for file transfer, but an extension towards other protocols such as SSH or FTP/TLS is foreseen. Moreover, SFelix supports the update of the meta-data of the bundle repository. These meta-data are a specific file that contains a description of bundles that are available on a given (or even several) bundle repositories. They are used by client to discover which bundles are available for download and installation, and to install them together with other bundles that are required for dependency resolution.

SFelix covers the client-side part of the deployment process. It validates all existing bundles at the platform launch time. Only valid bundles are installed, other one are ignored. In the case of the installation of new bundles, these latter are checked before their installation. This is done independently of the location of the bundle, being stored locally or retrieved from a bundle repository. Valid bundles are installed, unvalid ones rejected. During bundle update, the same verification occurs.

SFelix bundle validation occurs independently of the type of deployment trigger. It supports push deployment (initiated from the platform) as well as pull deployment (initiated from a remote shell or console).

  \subsubsection{Structure of the Program}

The general architecture of SFelix is the following. The security layer (which includes the bundle validation facility) is provided as a library used by the OSGi$^{tm}$ platform. This one has been slightly modified so as to check the validity of the signature of a bundle before installing it. The SFelix signer tool is provided as OSGi$^{tm}$ bundles.

\begin{figure}[htb]
\centering
\epsfig{file=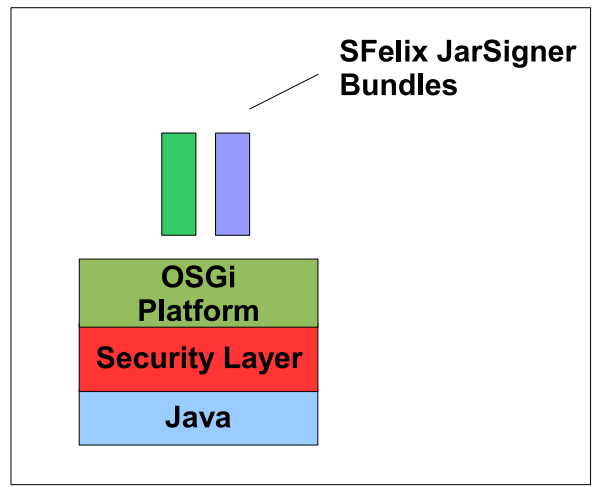, width=220pt}
\caption{The general Architecture of the Secure Felix (SFelix) Platform.}
\label{fig:secureFelix}
\end{figure}

Figure \ref{fig:secureFelix} shows the general architecture of the Secure Felix (SFelix) platform.

  \subsubsection{The API}

The API of SFelix is really simple. It is made of a method for signing bundles, and another one for validating their signature. 

The signature API is provided by the class fr.inria.ares.jarsigner.JarSigner. The method is named sign(), and takes as parameter the Jar file that is to be signed, the name of the file where the signed bundle is to be stored, as well as the name of the signer, the password to access the Keystore, and the password that protects the private key of the signer.

The signature validation API is provided by the class fr.inria.ares.jarvalidation.JarValidation. The single method is named check(), and takes as parameters the bundle to be verified (as a File object) and the password of the Keystore.

\begin{figure}[htb]
\centering
\epsfig{file=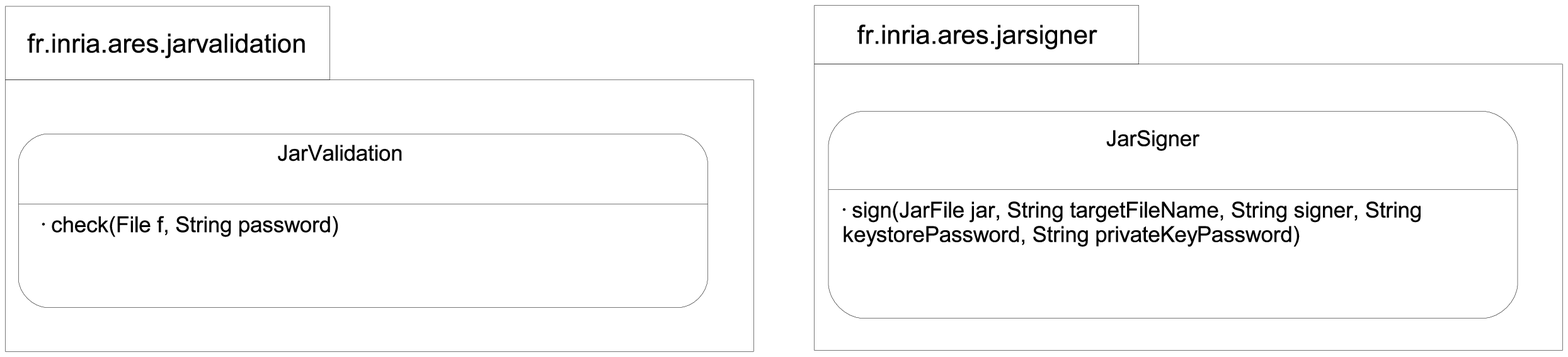, width=360pt}
\caption{The public API of the Secure Felix (SFelix) Platform.}
\label{fig:secureFelixApi}
\end{figure}

Figure \ref{fig:secureFelixApi} shows the public Application Programming Interface (API) of the SFelix platform. 

Next subsection will present with more detail the SFelix bundle signature validation facility, as well as modification that have been done to Felix to integrate this additional step on the deployment process.

 \subsection{Felix Modifications for Bundle Validation}

So as to support bundle validation, the algorithm presented in subsection \ref{detailedProcess} must be implemented, and executed for each bundle that is installed (or updated) on the platform. Moreover, the Felix platform must be slightly modified so as to integrated the stage of bundle validation in the installation process. These modifications build the bridge between the Felix and SFelix platforms. Modifications to the code will be presented, as well as the execution process at launch time and at runtime.

  \subsubsection{The Code}

The validation API, JarValidation, is provided as a separated library that is loaded at the launch time of the OSGi$^{tm}$ platform. This library is called immediately before the effective installation of each bundle. When the bundle is valid, installation is processed normally. In the contrary case, the installation aborts and the bundle is removed from the list of available bundles. All following bundles are checked and installed according to the same process.

The integration of the bundle validation process only requires the modification of three classes, BundleArchive, BundleCache and Felix, and the addition of the DefaultSecureBundleArchive class. All remaining code is provided in a separate library, jarvalidation.jar, which is required by the SFelix platform at launch time.

  \subsubsection{Launch Time of the SFelix Platform}

The SFelix platform aims at preventing malicious bundles to be installed and executed. It needs to achieve its goal while limiting as much as possible the interaction with its user (or manager). The security mechanisms must be as transparent as possible, so as to avoid deterring the users from exploiting them. For guaranteeing that the installed bundles are valid, it is necessary to be able to assert that the platform itself has not been tampered with. The validation process therefore occurs in two steps. First, the platform code must be verified. Secondly, the platform, that is known to be valid, checks each installed bundles.

 The validation of the platform code itself can be done manually through archive signing in a way similar to the bundle validation. However, in our case, the integrity of the platform code is simply verified through its hash value. The launch script containing original hash values is publicly available online on the project web site. Its execution guarantees the validity of the code archive.

The platform can then safely check the validity of the signature of external bundles. At launch time, all bundles are validated before their installation. If one bundle is not valid, it is simply rejected, and the installation of the other bundles goes on. For each bundle that is correctly installed, a confirmation of installation is printed in the shell to the user.

The only interaction between the platform and the user occurs through the (S)Felix shell during the validation of the first bundle. The user is asked for the password of the Keystore, which is necessary to retrieve the list of certificate that are considered trustful. Afterwards, the password is stored in the core of the platform, and reused for each validation of a bundle. Since the components only have access to the platform through the bundle context, and not directly, this way of storing the password is sound.

\begin{figure}[p]
\centering
\epsfig{file=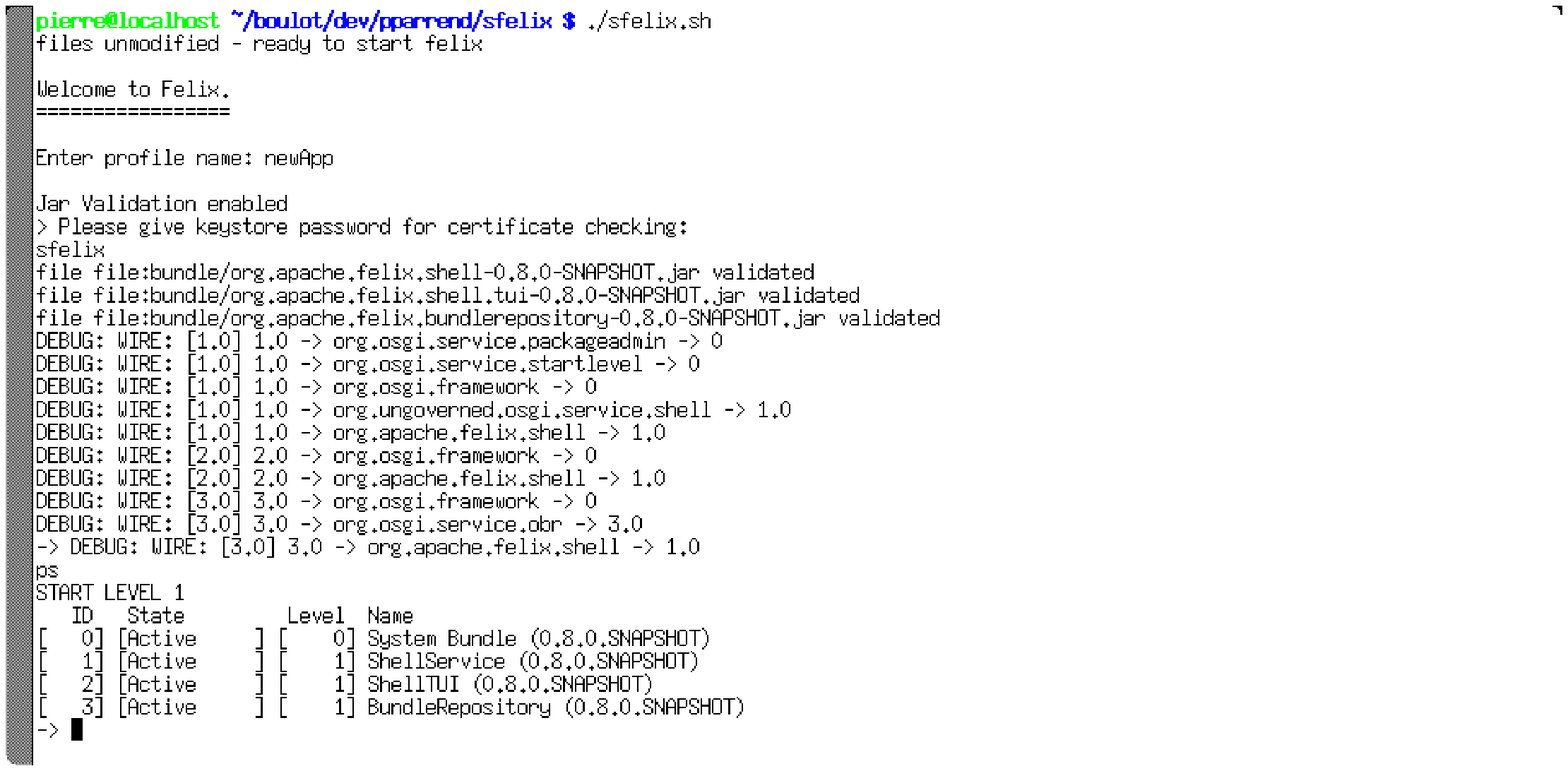, width=380pt}
\caption{Screen-shot of SFelix shell when launching a new SFelix Profile}
\label{fig:newApp}
-\end{figure}

The following code (Figure \ref{fig:newApp}) show the output when launching a new OSGi$^{tm}$ profile with SFelix. Note the password request and the notification of bundle validation.

  \subsubsection{Runtime of the SFelix Platform}

When bundles are installed during the runtime of the platform, or when bundles are updated, the same verification process occurs. Valid bundles are installed, and invalid one rejected. This is of course true independently of the location of installed bundles, whish can be local or stored in a remote repository.

\begin{figure}[p]
\centering
\epsfig{file=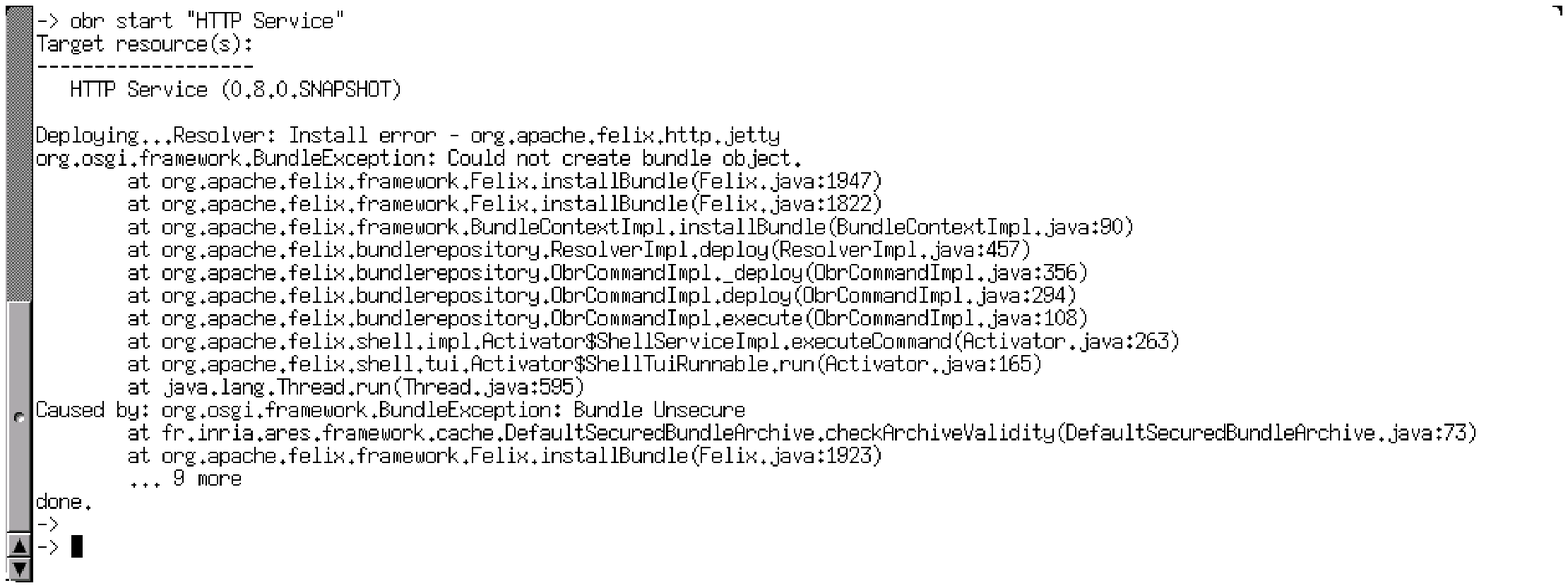, width=380pt}
\caption{Screen-shot of SFelix shell when trying to install an unsigned bundle}
\label{fig:BundleUnsecure-exception}
\end{figure}

Figure \ref{fig:BundleUnsecure-exception} shows a screen-shot of the Felix shell when trying to install an unsigned bundle.

\bigskip

The SFelix platform enables to validate all bundles that are executed before their installation. The correctness of the validation process is guaranteed by the verification of the platform code before launching the platform. In our current implementation, this is simply achieved through hash-value based verification of the code. More complete solutions are necessary to achieve high level security, but depends on the execution context of the platform.

The existence of a secure OSGi$^{tm}$ platform that validates bundles before installing them makes it necessary to have a tool available that support the process of signing them. We therefore developed SFelix JarSigner. So as to support the whole deployment process, an additional facility is included that enables to publish signed bundles in a remote file repository.

 \subsection{Tool: SFelix JarSigner}

The SFelix JarSigner tool aims at supporting the publication part of the process of component deployment. The publication is made of the signing of the bundles, and of their transfer to a public bundle repository. The functions of SFelix JarSigner are first presented, and the various bundles that compose it are detailed.

  \subsubsection{Functions of JarSigner}

JarSigner Graphical User Interface is composed of three main parts. The first aims at the connection to the Keystore. The second deals with bundle signing. The third is dedicated to the publication of bundles onto a remote repository.

\begin{itemize}
 \item  \textbf{The Keystore access} takes as input the name (Alias) of the person that wants to sign bundles. It also takes the general password of the Keystore, that enables to access the list of trusted certificates, and the password that protects the private key of the signer. The 'Open Session' button makes it possible to check the particular algorithm that is bound to the current alias.

 \item \textbf{The file signing part} enables to specify the name of the bundle that is to be signed, as well as the name of the future signed bundle. Note that these names must be different. Several actions over the bundle can be realized. It can of course be signed, but it can also simply be checked. When signing or checking a bundle through the 'Treat Bundle' button, the output of the process (success/failure) is printed in a specific information field at the bottom of the window. Moreover, bundles that are signed correctly or which signature is validated are added to the 'Selected Bundle' list, that makes it possible to chose which bundle is to be published.

 \item \textbf{The publication facility} of SFelix contains the above mentioned 'Selected Bundle' list, a list of file servers, and a 'Load File(s)' button that triggers the publication. Available bundles are exclusively the signed ones, but file servers can be added, modified and removed by the user of SFelix JarSigner.
\end{itemize}

\begin{figure}[p]
\centering
\epsfig{file=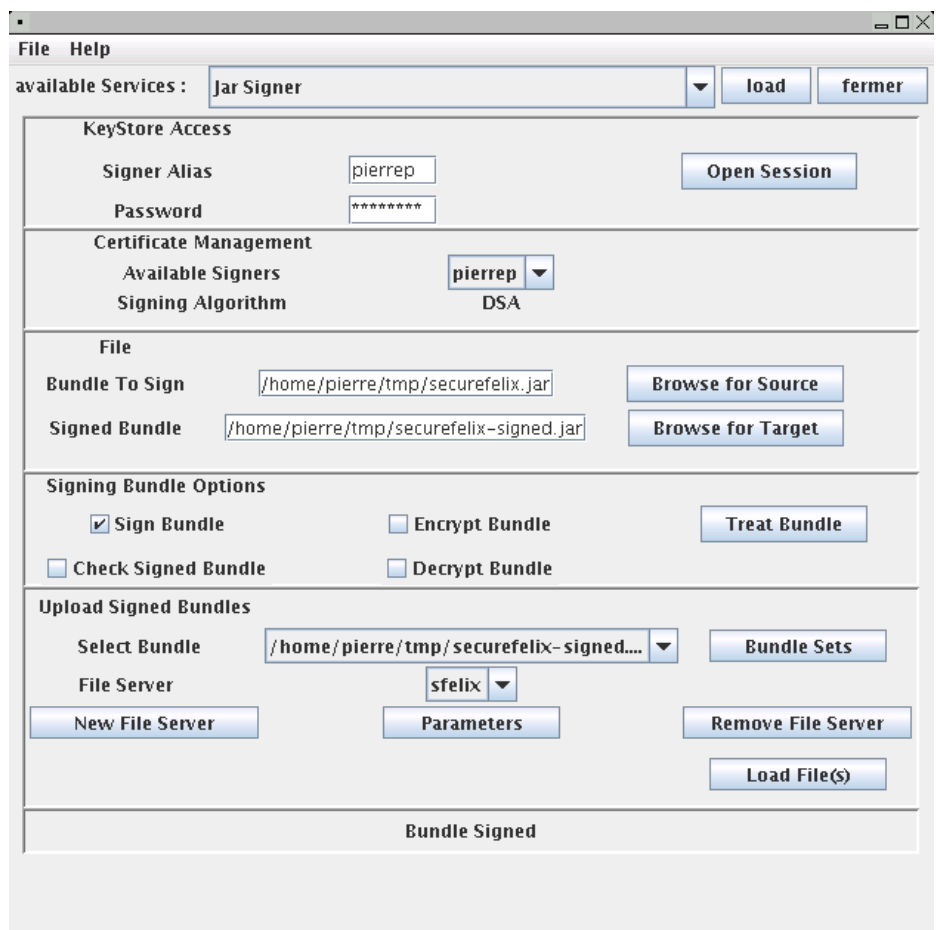, width=350pt}
\caption{The Graphical User Interface of the SFelix JarSigner Tool.}
\label{fig:jarSignerInterface-3}
\end{figure}

Figure \ref{fig:jarSignerInterface-3} shows the Graphical User Interface of the SFelix JarSigner tool. The different parts of the tool that have been presented in this section can be observed.

The modular organization of SFelix JarSigner will then be presented. 

  \subsubsection{Bundles of JarSigner}

SFelix JarSigner is an OSGi$^{tm}$ application. It is a tool that makes it possible to secure bundle deployment, and is itself made up of validated bundles: it is executed in the SFelix platform. To execute it in another OSGi$^{tm}$ platform would require to make the jarvalidation library available as an archive. This is quite easy to achieve, but, since it is not compliant with OSGi$^{tm}$ specifications, it is out of the scope of this study.

SFelix is built of three sets of code. The first one is the jarvalidation library. The second one is a lightweight plug-in support we developed for graphical interfaces named componentGui. The last one is of course the JarSigner tool itself.

The jarvalidation library provides the cryptographic library, the library for accessing to the Keystore, as well as the bundle validation API which is also used in JarSigner.

The componentGui facility is provided as a set of two bundles. The first one is named `SFelix Utilities', and provides various libraries for graphical interfaces elements. The second one is named `Generic Frame', and provides a simple graphical window that contains the list of all Graphical User Interfaces that are available in the platform. These GUIs are tagged by the fr.inria.ares.sfelix.utils.GuiSwingComponent programming interface they implement. They are made available as OSGi$^{tm}$ services.

The JarSigner tool is composed of two bundles, `Jar Signer', which provides the service for signing bundles, and `Jar Signer GUI', which provides the graphical interface that allows to access the signature service. This interface have been presented in the previous subsection.

\begin{figure}[p]
\centering
\epsfig{file=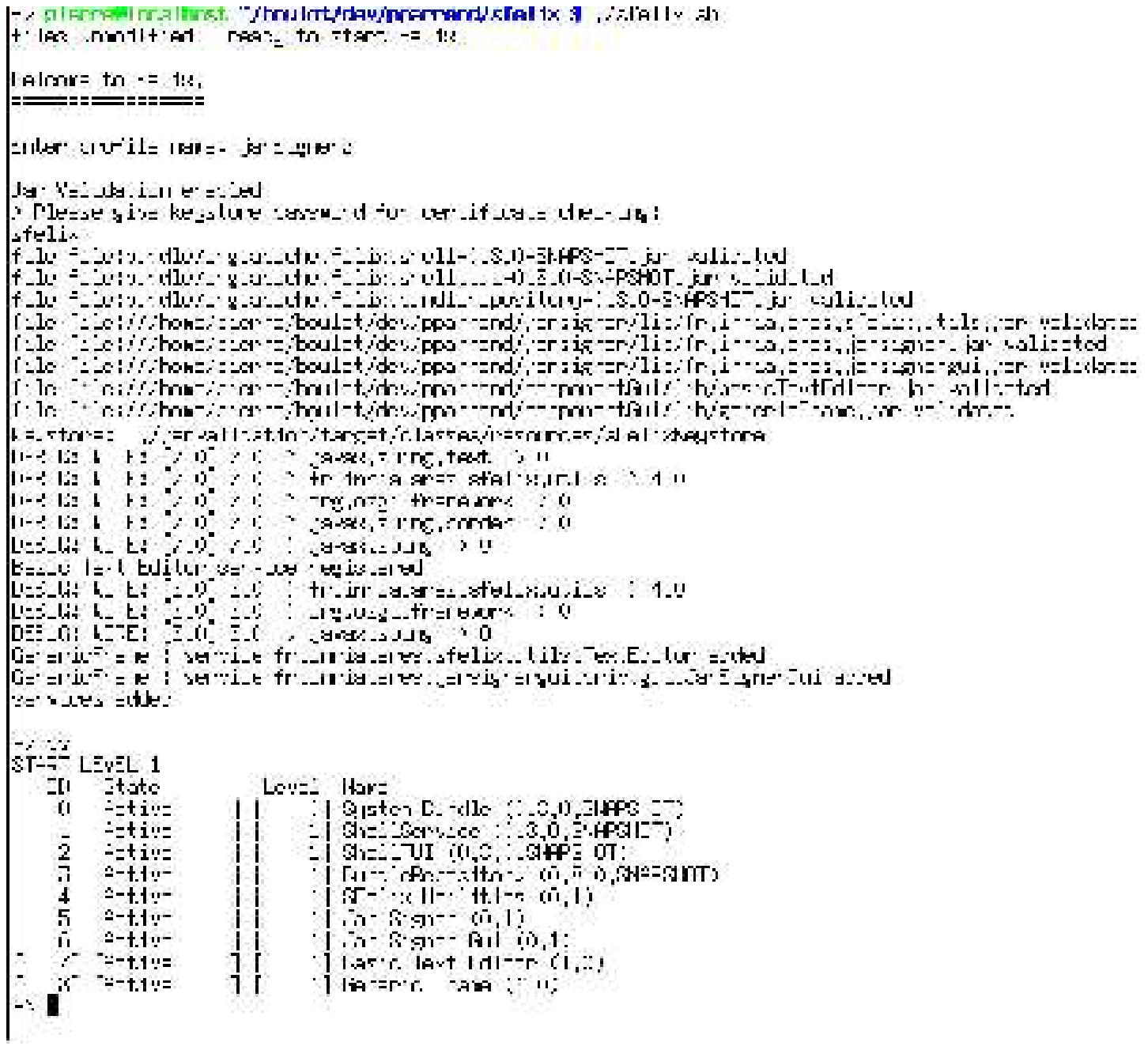, width=380pt}
\caption{Screen-shot of SFelix shell when launching the SFelix JarSigner Tool}
\label{fig:sfelix}
\end{figure}

The following code (figure \ref{fig:sfelix}) shows the output when launching SFelix JarSigner. The confirmation of the validation of the signature is printed for each bundle, and the various bundles that were presented are listed.

 \subsection{Conclusions}

In this section, the SFelix platform and the SFelix JarSigner tool have been presented. Used together, they support the whole process of secure deployment for OSGi$^{tm}$ bundles: signature, publication, and remote installation through the OBR 2 bundle repository if the bundles have a valid signature. SFelix is based on the Felix OSGi$^{tm}$ implementation: all bundles that run in SFelix also run on Felix, but Felix bundles need to be signed by a known person before being integrated in the secure SFelix platform.

%% file: conclusions.tex
\section{Conclusions}
\label{conclusions}

Until now, only very few effort seems to have been dedicated to securing component platforms. Due to the even broader dissemination of such platforms, it appears to be necessary to foster knowledge about security issues in component platforms. This work intends to make it possible for not security specialists to take such stakes into account when building a system based on a component platform. It is targeted to the OSGi$^{tm}$ platform, but presented concepts can easily be mapped towards other components systems.

This technical report gives a detailed overview of mechanisms that intervene during component signature and validation, including the cryptographic concepts that are necessary to understand the whole process.

Matching implementation of bundle signature and validation is introduced. Bundle validation is part of OSGi$^{tm}$ Release 4 Security Layer, and is as such integrated in the OSGi$^{tm}$ framework. Our implementation is available in the SFelix framework,which is an extension of the Felix OSGi$^{tm}$ implementation. Bundle signature is provided as a stand-alone application, SFelix JarSigner. This tool also supports publication of bundle in publicly accessible servers.

This work has brought to light further needs for security in component platforms. In particular, bundle validation implies that clients hava reliable informations about the signer. Several questions emerge: how to make sure clients have access to all bundle they are allowed to install~? How to restrict the access from certain clients to certain signers~? How to deal with new signers~? And how to deal with previous signers that are no longer allowed to publish bundles~? Moreover, it can sometime be necessary to be able to revoke isolated bundles, without preventing valid bundles to be installed.

%% file: annexes.tex
\section{Annexes}

 \subsection{Existing Tools for Secure Java Applications}

Existing tools that are add-ons to the Java Virtual Machine are jar signing facilities, and class ciphering software.

\paragraph{Signing Jars}

Some facilities already exist for signing jar. The main one is Sun Jarsigner \cite{jarSignerManual}, that provides a command line utility for signing jars. It resembles much to OSGi Security Layer, but has a different approach:
\begin{itemize}
 \item it is no Java program, but a command line tool. This is not consistent with OSGi specification, which states that the Security Layer is placed between the Java Virtual Machine and OSGi platform,
 \item one can not assume that an OSGi platform executing in a previsouly unknown environnement have necessary rights to execute third party program through JNI, nor that this third party program (here Sun Jarsigner) is available
 \item and last but not least, OSGi specification brings its own contraints on archive signature that are not specified by Jar specification, and thus not enforced by Sun Jarsigner. 
\end{itemize}

Moreover, no readily available library for signing jar is available. One implementation has been proposed by Raffi Krikorian in an On-Java article. However, this implementation use a Sun API that is no longer supported, as far as it has been proved to be insecure.

\paragraph{Ciphering of Classes} 

Besides signing archive, an other way to protect classes is to encrypt them, and to decrypt them only at runtime for the execution. This technique enables to guarantee not only integrity of sources and authentication of the emitter, but also confidentiality against potential malicious third parties.

All such libraries that are available are not free. Two of them, Canner and Katirya, are currently only available for the Microsoft Windows environment.
Canner, by Cinnabar Systems\footnote{http://www.cinnabarsystems.com/canner.html}, creates an executable file that then executes on the local JVM.
Katirya\footnote{http://www.mycgiserver.com/$\sim$ipnetdevelop/katirya.html} works according to the same principle.
jLock is the only tool that do not only work on MS Windows. It patches the Virtual Machine so as to integrate runtime decryption of encrypted classes\footnote{http://www.jbitsoftware.com/JBit/do/displayPage?targetPageId=products.jlockinfo}.

Available tools for ensuring security in Java applications are still quite limited. This is explained by the fact that most security problems are application and environnement specific. Therefore, effort for improving security in java system is either centered on the Virtual Machine itself - which does its job in a quite satisfactory manner - or on providing tool sets for specific applications. It is thus necessary to developp our own tools for implementing OSGi Security Layer.

 \subsection{Java Cryptographic Libraries}

Developping our own tools for enforcing Java security means providing a generic security API for signing and verifying jars. Such an API needs to releave on a valid implementation of cryptographic algorithms for hash value, digital signature and encryption. Such an implementation is of course out of scope of our work, and several cryptographic libraries have been developped recently.

This libraries can easily be plugged in java programs through the provider mechanism: by indicating the abbreviation matching an available provider at a cryptographic method call, the caller ensures that this provider is the one that performs the cryptographic operation. This makes it possible to integrate third party providers, but also to validate them independantly of their applications.

A list of security provider is maintained by Sun \footnote{http://java.sun.com/ products/jce/jce122\_providers.html}.
The principal open source cryptographic library has been developped by the Legion of the Bouncy Castle\footnote{http://www.bouncycastle.org/}.
However, for applications that need more than a basic level of security, these cryptographic libraries need to be validated. The reference organization for cryptographic module validation is the American National Institute of Standards and Technology (NIST) that has developped the Federal Information Processing Standards (FIPS) program for validating cryptographic libraries \cite{nistValidationProg}. Several Java libraries have undergone such a validation, some of them are available for use as independant software parts. These libraries are RSA BSAFE Crypto-J, IBM SSLite and CryptoLite, Certicom Security Builder FIPS Java Module, and Entrust Authority Security Toolkit for the Java Platform.

 \subsection {Algorithm for Bundle Signature and Validation}
\label{annexe:algos}

\begin{figure}[p]
\centering
\epsfig{file=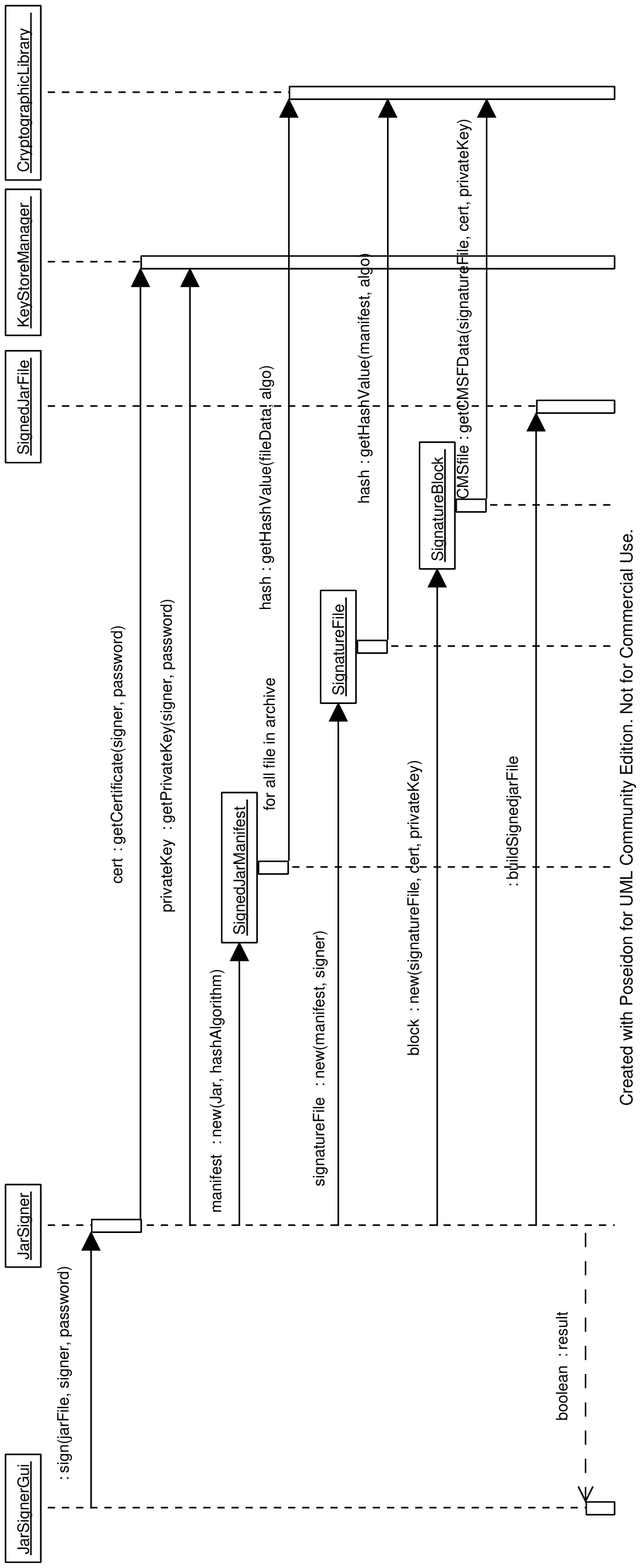, width=260pt}
\caption{The Sequence Diagram of the Algorithm for signing a Bundle.}
\label{fig:seq-signing}
\end{figure}

Figure \ref{fig:seq-signing} shows the detailed sequence diagram of the algorithm for signing a bundle.

\begin{figure}[p]
\centering
\epsfig{file=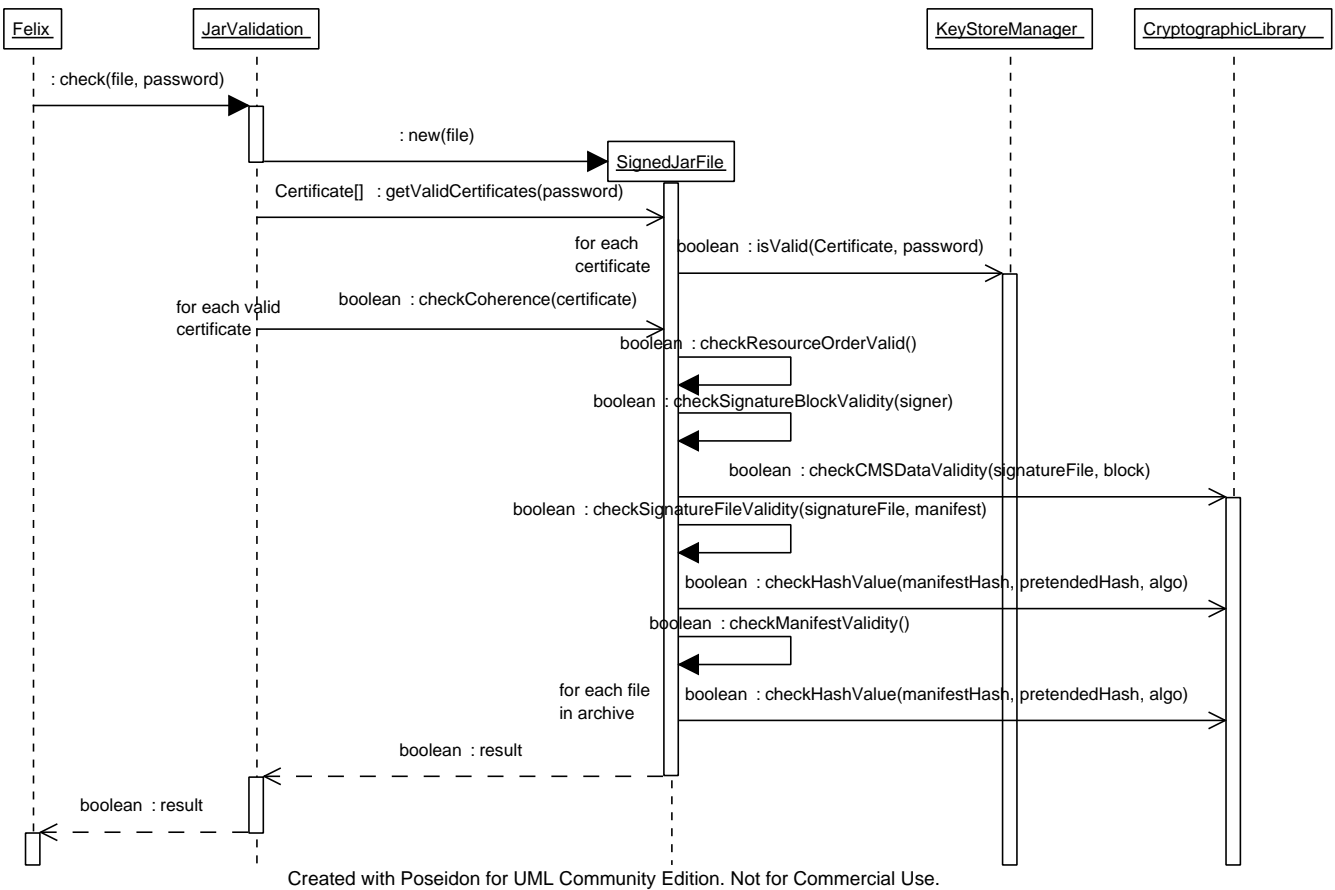, width=350pt}
\caption{The Sequence Diagram of the Algorithm for validating a signed Bundle.}
\label{fig:seq-verification}
\end{figure}

Figure \ref{fig:seq-verification} shows the detailed sequence diagram of the algorithm for validating a signed bundle.

